\documentclass[aps,pra,twocolumn,groupedaddress,showpacs]{revtex4}

\usepackage{graphicx}
\usepackage{amsmath}

\begin{document}


\title{Modulation instability and solitary wave formation in two-component Bose-Einstein condensates}


\author{Kenichi Kasamatsu$^{1}$}
\author{Makoto Tsubota$^{2}$}

\affiliation{$^1$Department of General Education, 
Ishikawa National College of Technology, Tsubata, Ishikawa 929-0392, Japan\\
$^2$Department of Physics,
Osaka City University, Sumiyoshi-Ku, Osaka 558-8585, Japan}


\date{\today}

\begin{abstract}
We investigate nonlinear dynamics induced by the modulation instability of a two-component mixture in an atomic Bose-Einstein condensate. The nonlinear dynamics is examined using numerical simulations of the time-dependent coupled Gross-Pitaevskii equations. The unstable modulation grows from initially miscible condensates into various types of vector solitary waves, depending on the combinations of the sign of the coupling constants (intracomponent and intercomponent). We discuss the detailed features of the modulation instability, dynamics of solitary wave formation, and an analogy with the collapsing dynamics in a single-component condensate with attractive interactions. 
\end{abstract}


\maketitle

\section{INTRODUCTION}
Dynamics of spatial pattern formation in nonlinear media is important for a wide range of physical phenomena \cite{Cross}. Modulation instability (MI) is an indispensable mechanism for understanding pattern formation from a uniform medium. MI occurs when a constant-wave background becomes unstable to induced sinusoidal modulations under the combined effects of nonlinearity and diffraction in a spatially nonlinear field. Also, it is known that MI causes a uniform medium to break-up into pulsed ``solitary waves" \cite{Agrawalbook}. The effect of self-interactions in nonlinear media plays a crucial role for the MI in a single component system. In a multicomponent system, in which there is more than one order parameter, additional types of interactions can occur between different components and have a great influence on the MI. 

Atomic-gas Bose-Einstein condensates (BECs) are a good system for examining MI and nonlinear matter-wave dynamics. In this system, the nonlinearity originates from the atom-atom interaction. Moreover, manipulation of the matter waves can be achieved by applying established techniques from atomic, molecular, and optical physics. For example, in single-component BECs, researchers have studied nonlinear excitations such as dark solitons \cite{Burger,Denschlag,Anderson}, bright solitons \cite{Khaykovich,Strecker,Eiermann}, and quantized vortices \cite{Matthews,Madison}. It has been clarified that the MI plays a crucial role in the formation process of those excitations and gives rise to the intriguing nonlinear dynamics \cite{Salasnich,Kamchatnov,Carr,Konotop,Smerzi}. 

This paper addresses the nonlinear dynamics of solitary wave formation induced by MI in a mixture of two-component atomic-gas BECs. The experimental creation of multicomponent BECs has been achieved by the simultaneous magnetic trapping of the atoms lying in the week-field-seeking states \cite{Hall,Maddaloni}, the use of an optical trapping that liberates the spin degrees of freedom of atoms \cite{Stenger,Barrett} or simultaneous trapping of different species of atoms \cite{Modugno,Mudrich,Modugno2}. The MI in two-component BECs was firstly discussed by Goldstein and Meystre \cite{Goldstein}, and recently reexamined \cite{Raju}. However, how the condensates develop under the MI is still unclear, though some progress can be seen for the study of spin-1 BECs \cite{spin1}. 

First, we summarize the MI condition with respect to two intracomponent interactions and an intercomponent one. Although all of these atom-atom interactions were repulsive in the past experiments of multicomponent BECs, except for a boson-fermion mixture in Ref. \cite{Modugno2}, further insight can be gained if we consider some of them are attractive. The character of the interactions may be controlled by choosing the specific kinds of atom \cite{Modugno,Mudrich,Modugno2} or by using a homonuclear \cite{Khaykovich,Strecker} or heteronuclear Feshbach resonance \cite{Erhard,Stan}. Next, we discuss the nonlinear dynamics caused by the MI, emphasizing the role of the intercomponent interaction. We focus on the situation in which one component is suddenly put on the other component with repulsive interaction. The MI will lead to the formation of a {\it vector soliton} train in such a way that a bright soliton train is generated through the MI in a single-component condensate \cite{Salasnich,Kamchatnov,Carr}. Our previous paper \cite{Kasamatsu} revealed the dynamics of domain formation of two-component BECs in the case of all interactions being repulsive, in excellently agreement with the experimental observation of Miesner {\it et al.} \cite{Miesner}. We extend the analysis to the cases where two components have attractive intercomponent interaction, or one of them has an attractive intracomponent interaction. Depending on the combination of the sign of the s-wave scattering lengths, two-component BECs exhibit rich nonlinear dynamics of solitary wave formation. 

This paper is organized as follows. In Sec. \ref{modell}, we formulate of the problem for two-component BECs using a quasi-one-dimensional model for simplicity. Then we use linear stability analysis to clarify the MI with respect to the sign of the coupling constants. Section \ref{numeri} presents the numerical simulation results that confirm the MI analysis and show how the solitary wave formation occurs in the condensates through the MI. Section \ref{condle} is devoted to conclusion. 

\section{MODULATION INSTABILITY OF TWO-COMPONENT BOSE EINSTEIN CONDENSATES} \label{modell}
\subsection{Model}
We start with a two-component BEC with atomic masses $m_{1}$ and $m_{2}$. The dynamics can be derived by assuming that the two condensates are described by the wave functions $\Psi_{1}({\bf r},t)$ and $\Psi_{2} ({\bf r},t)$. At zero temperature, the total energy functional of the system is 
\begin{eqnarray}
E[\Psi_{1},\Psi_{2}] = \int d {\bf r} \biggl[ \frac{\hbar^{2}}{2m_{1}} |\nabla \Psi_{1}|^{2} + \frac{\hbar^{2}}{2m_{2}} |\nabla \Psi_{2}|^{2} \nonumber \\ 
+ V_{\rm ext}^{(1)} |\Psi_{1}|^{2} + V_{\rm ext}^{(2)} |\Psi_{2}|^{2} + \frac{1}{2} g_{1} |\Psi_{1}|^{4} + \frac{1}{2} g_{2} |\Psi_{2}|^{4} \nonumber \\ 
+ g_{12} |\Psi_{1}|^{2} |\Psi_{2}|^{2} \biggr]. 
\end{eqnarray}
The condensates are assumed to be trapped in axisymmetric harmonic potentials: 
\begin{equation}
V_{\rm ext}^{(i)}(r,z) = \frac{1}{2} m_{i} \omega_{i}^{2} (r^{2} + \lambda^{2} z^{2}),
\hspace{3mm}  i=1,2, 
\end{equation}
where $\omega_{i}$ is the transverse trapping frequency and $\lambda$ is the aspect ratio of the potential. Each component can have its  own trapping frequency due to the $g$-factor and index of the atomic hyperfine levels along the quantized axis. The intracomponent coupling constant $g_{i}=4 \pi \hbar^{2} a_{i}/m_{i}$ is characterized by the scattering lengths $a_{1}$ and $a_{2}$ between atoms of the same species, while the intercomponent one $g_{12} = 4\pi\hbar^{2}a_{12}/m_{12}$ ($m_{12}^{-1}=m_{1}^{-1}+m_{2}^{-1}$) is determined by the scattering length $a_{12}$ where an atom in the $\Psi_{1}$ component scatters from another atom in the $\Psi_{2}$ component. This intercomponent coupling yields new structures and dynamics not found in a single component BEC \cite{Hall,Maddaloni,Modugno,Goldstein,Raju,Kasamatsu,Miesner,Ho,Gordon,Busch,Oehberg}.

The dynamics of two-component BECs can be described using the coupled GP equations, which are derived from the variational principle $i \hbar \partial \Psi_{i}/\partial t = \delta E/\delta \Psi_{i}^{\ast}$ as 
\begin{subequations}
\begin{eqnarray}
i\hbar \frac{\partial \Psi_{1}}{\partial t} = \biggl( -\frac{\hbar^{2} \nabla^{2}}{2m_{1}}
+V_{\rm ext}^{(1)} + g_{1} |\Psi_{1}|^{2} + g_{12} |\Psi_{2}|^{2} \biggr) \Psi_{1}, \\
i\hbar \frac{\partial \Psi_{2}}{\partial t} = \biggl( -\frac{\hbar^{2} \nabla^{2}}{2m_{2}}
+V_{\rm ext}^{(2)} + g_{2} |\Psi_{2}|^{2} + g_{12} |\Psi_{1}|^{2} \biggr) \Psi_{2}. 
\end{eqnarray} \label{2tgpe}
\end{subequations}
The normalization of each wave function is taken independently as $\int d{\bf r} |\Psi_{i}({\bf r})|^{2}=N_{i}$. 

We assume that the condensates are tightly confined in the transverse direction, so $\lambda \ll 1$. This condition means that the motional degrees of freedom in the $x$-$y$ plane are frozen, a situation that could be realized in highly elongated cigar-shaped potentials. In this case, one can factorize the condensate wave function into a longitudinal and a transverse part as 
\begin{equation}
\Psi_{i}({\bf r},t) = \phi_{\perp}^{(i)} (x,y) \psi_{i} (z,t) e^{-i \omega_{i} t},
\end{equation}
where $\phi_{\perp}^{(i)} (x,y)$ is the normalized ground state of the transverse potential $V_{\rm ext}^{(i)}(r)=m_{i}\omega_{i}^{2} r^{2} / 2$ with energy $\hbar \omega_{i}$. The system is thus effectively reduced to a one-dimensional geometry, with the longitudinal condensate wave function $\psi_{i} (z,t)$ satisfying the one-dimensional GP equations: 
\begin{subequations}
\begin{eqnarray}
i\hbar \frac{\partial \psi_{1}}{\partial t} = \biggl( -\frac{\hbar^{2}}{2m_{1}} \frac{\partial^{2}}{\partial z^{2}} 
+\frac{1}{2}m_{1} \lambda^{2} \omega_{1}^{2}z^{2} + u_{1} |\psi_{1}|^{2} \nonumber \\ 
+ u_{12} |\psi_{2}|^{2} \biggr) \psi_{1}, \\
i\hbar \frac{\partial \psi_{2}}{\partial t} = \biggl( -\frac{\hbar^{2}}{2m_{2}} \frac{\partial^{2}}{\partial z^{2}} 
+\frac{1}{2}m_{2} \lambda^{2} \omega_{2}^{2} z^{2} + u_{2} |\psi_{2}|^{2} \nonumber \\ 
+ u_{12} |\psi_{1}|^{2} \biggr) \psi_{2}.
\end{eqnarray} \label{2tgpe1D}
\end{subequations}
Here, 
\begin{eqnarray}
u_{i} = g_{i} \eta_{i} = g_{i} \int dx dy |\phi_{\perp}^{(i)}|^{4} = \frac{g_{i}}{2 \pi b_{i}^{2}}, \\
u_{12} = g_{12} \eta_{12} =g_{12} \int dx dy |\phi_{\perp}^{(1)}|^{2} |\phi_{\perp}^{(2)}|^{2} = \frac{g_{12}}{\pi (b_{1}^{2} + b_{2}^{2})}
\end{eqnarray}
with the length scale $b_{i}=\sqrt{\hbar/m_{i} \omega_{i}}$ characteristic of $V_{\rm ext}^{(i)}(r)$.

\subsection{Relation with the nonlinear Schr\"{o}dinger equation in nonlinear optics}
 In the context of nonlinear optics, a special attention has been paid to MI in Kerr media in which light-wave propagation is described by the nonlinear Schr\"{o}dinger equation (NLSE) within the scalar approximation of the electromagnetic field \cite{Agrawalbook}. The NLSE exhibits instability of self-phase-modulation (SPM) when nonlinearity and group velocity dispersion (GVD) act in opposition, e.g., for self-focusing waves associated with negative nonlinearity the GVD should be ``normal" (a positive GVD coefficient) and for self-defocusing waves associated with positive nonlinearity the GVD should be ``anomalous" (a negative GVD coefficient). This condition is also necessary for the existence of bright solitons which result from an exact balance between nonlinearity and dispersion. 
 
If accounting for polarization of the electromagnetic field, light propagation in isotropic Kerr media is described by two incoherently coupled NLSEs instead of the single NLSE \cite{Agrawalbook}. Then, the incoherent coupling between two NLSEs, referred to as cross-phase-modulation (XPM), leads to MI for any sign of nonlinearity and GVD \cite{Agrawal}. The XPM is a general phenomenon characteristic of the simultaneous nonlinear propagation of several waves belonging to different modes. Also, MI induced by the XPM is of fundamental importance as it suggests the possibility of soliton formation in the normal dispersion regime. 

Equations (\ref{2tgpe1D}) have a close analogy with the incoherently coupled NLSEs in nonlinear optics, where the $u_{i}$- and $u_{12}$-terms correspond to the SPM and XPM terms, respectively. In nonlinear optics, the ratio of the nonlinear coefficients for SPM and XPM can be altered using the light's angle of elliptic polarization \cite{Agrawalbook}. For the atomic BECs, the strength of the atomic interactions can be altered using the Feshbach resonance \cite{Erhard,Stan}.  

\subsection{Modulation instability analysis}\label{moduinst}
In a single-component nonlinear wave, the MI induced by SPM exists only for the waves with self-focusing nonlinearity, corresponding to the attractive interaction between atoms. The intercomponent coupling (i.e., XPA) is a feature of the two-component system that does not exist in a single-component system. In this section, we discuss how the sign and strength of the coupling parameters $u_{1}$, $u_{2}$ and $u_{12}$ affect the MI. 

We examine the stability of miscible two-component BECs with the homogeneous one-dimensional density $n_{10}=|\psi_{10}|^{2}$ and $n_{20}=|\psi_{20}|^{2}$ \cite{Goldstein,Raju,Kasamatsu}. When the wave functions are written as $\psi_{i}(z, t)=\sqrt{n_{i0}} + \delta \psi_{i}(z, t)$, the linearized equation for the fluctuations becomes 
\begin{subequations}
\begin{eqnarray}
i \hbar \frac{\partial}{\partial t} \delta \psi_{1} = 
- \frac{\hbar^{2}}{2m_{1}} \frac{d^{2}}{dz^{2}} \delta \psi_{1} 
+u_{1} n_{10}(\delta \psi_{1}+\delta \psi_{1}^{\ast}) \nonumber \\
+u_{12} \sqrt{n_{10}n_{20}}(\delta \psi_{2}+\delta \psi_{2}^{\ast})  \\
i \hbar \frac{\partial}{\partial t} \delta \psi_{2} = 
- \frac{\hbar^{2}}{2m_{2}} \frac{d^{2}}{dz^{2}} \delta \psi_{2} 
+u_{2} n_{20}(\delta \psi_{2}+\delta \psi_{2}^{\ast}) \nonumber \\
+u_{12} \sqrt{n_{10}n_{20}}(\delta \psi_{1}+\delta \psi_{1}^{\ast})  
\end{eqnarray}\label{lineareq}
\end{subequations}
We assume a general solution of the form $\delta \psi_{i} = \zeta_{i} \cos (k_{i}z-\Omega t) + i \eta_{i} \sin (k_{i}z-\Omega t)$, where we allow for different wave numbers $k_{i}$ for the $\psi_{i}$ ($i=1,2$) components. Then, Eqs. (\ref{lineareq}) provide a set of equations for the amplitude $\zeta_{i}$ and $\eta_{i}$. Straightforward calculation gives the dispersion relation 
\begin{equation}
(\Omega^{2} - \Lambda_{1}) (\Omega^{2} - \Lambda_{2}) = P^{2},
\end{equation} 
where
\begin{eqnarray}
\Lambda_{i} = \frac{k_{i}^{2}}{2m_{i}} \left( \frac{\hbar^{2} k_{i}^{2}}{2m_{i}} + 2 u_{i} n_{i0} \right), \\
P = \frac{u_{12}}{\sqrt{m_{1}m_{2}}} \sqrt{n_{10} n_{20}} k_{1} k_{2}.
\end{eqnarray}
The dispersion relation gives a quadratic algebraic equation in terms of $\Omega^{2}$, whose solution is
\begin{equation}
\Omega_{\pm}^{2} = \frac{1}{2} \left[ \Lambda_{1}+\Lambda_{2} \pm \sqrt{ (\Lambda_{1}+\Lambda_{2})^{2} + 4 (P^{2} - \Lambda_{1} \Lambda_{2})  } \right]. \label{disersion}
\end{equation}

The condensates are uniformly miscible and their stability is governed by Eq. (\ref{disersion}). If the frequency $\Omega_{\pm}$ has an imaginary part, the spatially modulated perturbations grow exponentially with time, as is evident from the form of $\delta \psi_{i}$. This unstable growth of weak perturbations is referred to as the MI. The MI condition depends on the sign of two variables
\begin{eqnarray}
\Lambda \equiv \Lambda_{1} + \Lambda_{2}, \label{lambdacond} \\
\Delta \equiv P^{2} - \Lambda_{1} \Lambda_{2}; \label{Deltacond}
\end{eqnarray}
With these two variables, Eq. (\ref{disersion}) is rewritten as 
\begin{equation}
\Omega^{2}_{\pm} = \frac{1}{2} (\Lambda \pm \sqrt{\Lambda^{2} + 4\Delta}). \label{freqkika}
\end{equation}
For $\Lambda>0$, the value of $\Omega_{+}^{2}$ is always positive, whereas the $\Omega_{-}^{2}$ becomes negative only if $\Delta > 0$ in which case $\Omega_{-}$ is purely imaginary. For $\Lambda<0$, the value of $\Omega_{+}^{2}$ becomes negative when $\Delta < 0$ and $\Omega_{-}^{2}$ is always negative; thus, the system is always modulationally unstable. 

\subsection{Condition of MI for a single-component condensate}\label{singleMIrev}
Before considering the general case, it is instructive to review briefly the MI of a single-component BEC. For a single component, $u_{2}=0$, $u_{12}=0$ ($P=0$), and the dispersion relation Eq. (\ref{disersion}) reduces to $\Omega^{2}=\Lambda_{1}$. The MI occurs when $\Lambda_{1}=\hbar^{2} k_{1}^{2}/2 m_{1} + 2 u_{1} n_{10} < 0$. We obtain the well known result that the MI occurs only with an attractive coupling constant $u_{1}<0$. In this case, the imaginary component of the frequency $G={\rm Im} \Omega$ represents the growth rate of the modulation, which is called the gain spectra \cite{Agrawalbook}. This component is given by $G=\frac{k_{1}}{2m_{1}} \sqrt{4m_{1}|u_{1}|n_{10}-\hbar^{2} k_{1}^{2}}$ in the range $0 < k_{1} < \sqrt{4m_{1}|u_{1}|n_{10}}/\hbar$. The fastest growth occurs for the wave number $k_{1 \rm{max}}$ that gives a maximum of $G$. The extremum condition $\partial \Omega^{2}/\partial k_{1}^{2}=0$ gives $k_{1{\rm max}}=\sqrt{2m_{1}|u_{1}|n_{10}}/\hbar$ and the maximum growth rate $G_{\rm max}=|u_{1}| n_{10}/\hbar$. The MI associated with the attractive interaction has a key role in the formation of bright solitons of a single-component BEC \cite{Salasnich,Carr}.

\subsection{Condition of MI for a two-component condensate}
This study is concerned with the MI relevant to the intercomponent coupling; thus, we fix the interaction of the $\psi_{1}$-component to be positive $u_{1}>0$. Possible choices of the sign of the coupling strengths $u_{2}$ and $u_{12}$ are summarized in Table \ref{coupclas}. To classify the types of the instability more clearly, we introduce the length scale $\ell^{2} \equiv (4 m_{1} u_{1} n_{10}/\hbar^{2})^{-1}$ and the dimensionless wave number $\tilde{k}_{i}=k_{i} \ell$. Then, Eqs. (\ref{lambdacond}) and (\ref{Deltacond}) become
\begin{eqnarray}
\Lambda = \frac{\hbar^{2}}{4m_{1}^{2} \ell^{4}} \left[ \tilde{k}_{1}^{2} (\tilde{k}_{1}^{2} + 1) + \frac{m_{1}^{2}}{m_{2}^{2}} \tilde{k}_{2}^{2} (\tilde{k}_{2}^{2} + \gamma_{2}) \right], \label{selpmu} \\
\Delta = \left( \frac{\hbar^{2}}{4 m_{1}^{2} \ell^{4}} \right)^{2} \tilde{k}_{1}^{2} \tilde{k}_{2}^{2} \left[ \gamma_{12}^{2} - (\tilde{k}_{1}^{2} + 1)(\tilde{k}_{2}^{2} + \gamma_{2}) \right], \label{crosspmu}
\end{eqnarray}
where 
\begin{equation}
\gamma_{2} = \frac{m_{2} u_{2} n_{20}}{m_{1} u_{1} n_{10}}, \hspace{5mm}  \gamma_{12} =  \frac{u_{12}}{u_{1}} \sqrt{\frac{m_{2} n_{20}}{m_{1} n_{10}}}.
\end{equation}
These equations show that the MI condition depends on two atomic masses, condensate density, three coupling constants, and the range of the wave numbers. Here, we assume $m_{1}=m_{2}$, which greatly simplifies the form of the following equations. 
\begin{table}
\caption{\label{coupclas} Four cases that are considered, each defined by the sign (positive = ``$+$h, negative = ``$-$h) of the coupling constants $u_{2}$ and $u_{12}$. In all cases, $u_{1}$ is assumed to be positive.}
\begin{tabular}{cccccc}
 & $u_{2}(\gamma_{2})$ & $u_{12} (\gamma_{12})$ \\
\hline
(a) & $+$ & $+$ \\
(b) & $+$ & $-$ \\
(c) & $-$ & $+$ \\
(d) & $-$ & $-$  
\end{tabular}
\end{table}

We search the unstable region of the wave number $\tilde{k}_{i}$ by changing the values of $\gamma_{2}$ and $\gamma_{12}$. By examing the possible choices for the signs of $\gamma_{2}$ and $\gamma_{12}$ shown in Table \ref{coupclas}, we obtained the unstable region in $\tilde{k}_{1}$-$\tilde{k}_{2}$ space as shown in Fig. \ref{pararega} and \ref{pararegb}. We summarize below some features of the instability. 

\subsubsection{Region (a): $\gamma_{2}>0$ and $\gamma_{12}>0$}
\begin{figure}
\includegraphics[height=0.32\textheight]{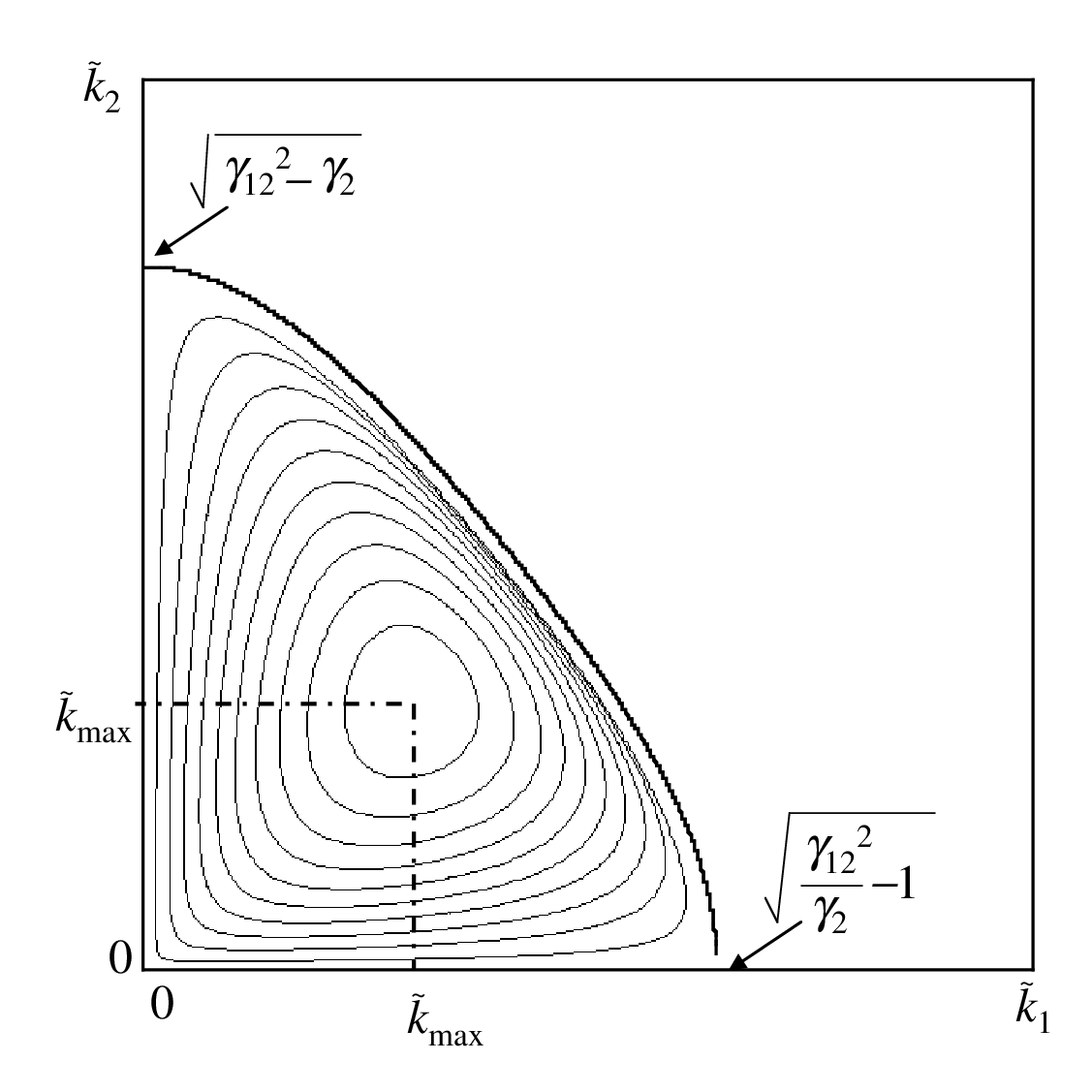}
\caption{The modulationally unstable region in $\tilde{k}_{1}$-$\tilde{k}_{2}$ space for the combinations (a) and (b) in Table \ref{coupclas}. The unstable region lies below the thick line boundary given by $\tilde{k}_{2} = \sqrt{\gamma_{12}^{2} /(\tilde{k}_{1}^{2}+1) - \gamma_{2}} $. The magnitude of the gain spectra $G={\rm Im} \Omega$ (arbitrary unit) is shown by a contour plot in the unstable region.}
\label{pararega}
\end{figure}
This situation is of particular importance because the individual condensates with the repulsive interaction are modulationally stable. In this case, $\Lambda>0$ and thus $\Omega_{+}^{2}$ is always positive. Therefore, the unstable condition is determined from Eq. (\ref{crosspmu}), where the positive $\Delta$ ($ \Delta > 0 $) gives a purely imaginary $\Omega_{-}$. For fixed $\tilde{k}_{2}$ the instability occurs within a certain range of $\tilde{k}_{1}$; $0<\tilde{k}_{1}<\sqrt{\gamma_{12}^{2}/(\tilde{k}_{2}^{2}+\gamma_{2})-1}$ as shown in Fig \ref{pararega}. For the wavenumber to be real, the term in the square root must be positive, which gives the necessary condition $\gamma_{12} > \sqrt{\gamma_{2}+\tilde{k}_{2}^{2}}$ or $\gamma_{12} < - \sqrt{\gamma_{2}+\tilde{k}_{2}^{2}}$ for the MI to occur. The former corresponds to the strong repulsive intercomponent interaction and the latter to the corresponding attractive interactions. Thus, the unstable range is independent of the sign of $\gamma_{12}$. In the former case, we obtain the well-known condition $\sqrt{g_{1}g_{2}}<g_{12}$ for phase separation in the long wavelength limit $k_{i} \rightarrow 0$ \cite{Ho}. In Fig. \ref{pararega}, we also show the magnitude of the imaginary component of $\Omega_{-}$ (gain spectra $G={\rm Im} \Omega_{-}$). The maximum of $G$ appears at the wave number $\tilde{k}_{1}=\tilde{k}_{2} = \tilde{k}_{\rm max}$. After setting $\tilde{k}_{1}=\tilde{k}_{2}=\tilde{k}$ in Eq. (\ref{freqkika}), the most unstable wave number is calculated from $\partial \Omega_{-}^{2}/\partial \tilde{k}^{2} = 0$, with the result 
\begin{equation}
\tilde{k}_{\rm max} = \frac{1}{2} \left( \sqrt{(\gamma_{2}-1)^{2} + 4 \gamma_{12}^{2}} - \gamma_{2} - 1 \right)^{1/2} \label{maxwavenum}
\end{equation}
and the maximum growth rate becomes 
\begin{equation}
G_{\rm max} = \frac{\hbar \tilde{k}_{\rm max}^{2}}{ 2 m \ell^{2} }=\frac{\hbar}{8 m \ell^{2}} \left( \sqrt{(\gamma_{2}-1)^{2} + 4 \gamma_{12}^{2}} - \gamma_{2} - 1 \right). \label{maxgrowth}
\end{equation}

The unstable modulation develops by following the eigenvectors associated with the eigenvalue $\Omega_{-}$. For $\tilde{k}_{1}=\tilde{k}_{2}=\tilde{k}$, a simple calculation gives the mode amplitudes as 
\begin{eqnarray}
\left(
\begin{array}{c}
\zeta_{1 \pm} \\
\zeta_{2 \pm} 
\end{array}
\right) = \left(
\begin{array}{c} 
\frac{1-\gamma_{2}}{2 \gamma_{12}} \pm {\rm sign}(u_{12}) \sqrt{1 + \frac{(1-\gamma_{2})^{2}}{4 \gamma_{12}^{2}}} \\ 
1
\end{array} 
\right), \label{eigenvecmatrix} \\
\eta_{i \pm}=\frac{2 m \Omega_{-} }{ \hbar k_{i}^{2} } \zeta_{i \pm}   \hspace{5mm} i=1,2.
\end{eqnarray}
The positive (negative) sign represents the eigenvector associated with $\Omega_{+}$ ($\Omega_{-}$). For positive $u_{12}$ the amplitude $\zeta_{1+}$ ($\zeta_{1-}$) is always positive (negative), which means that the unstable modulation $\zeta_{i-}$ develops into out-of-phase waves. This feature follows from the fact that the repulsive character of the intercomponent interaction forces the two components apart.

\subsubsection{Region (b): $\gamma_{2}>0$ and $\gamma_{12}<0$}
In this case, although each component has a repulsive intracomponent interaction, the two components have a strong attraction. Because the combination of the intercomponent coupling is included through $\gamma_{12}$, this case is similar to that of case (a). Thus, the most unstable wave number and the corresponding maximum gain spectra are the same as the combination (a) with only the signs of the modulation amplitudes being different: $\zeta_{i+}<0$ and $\zeta_{i-}>0$ in Eq. (\ref{eigenvecmatrix}). Therefore, the MI leads to an in-phase evolution of the two-component modulation. 

\subsubsection{Region (c): $\gamma_{2}<0$ and $\gamma_{12}>0$}
\begin{figure}
\includegraphics[height=0.32\textheight]{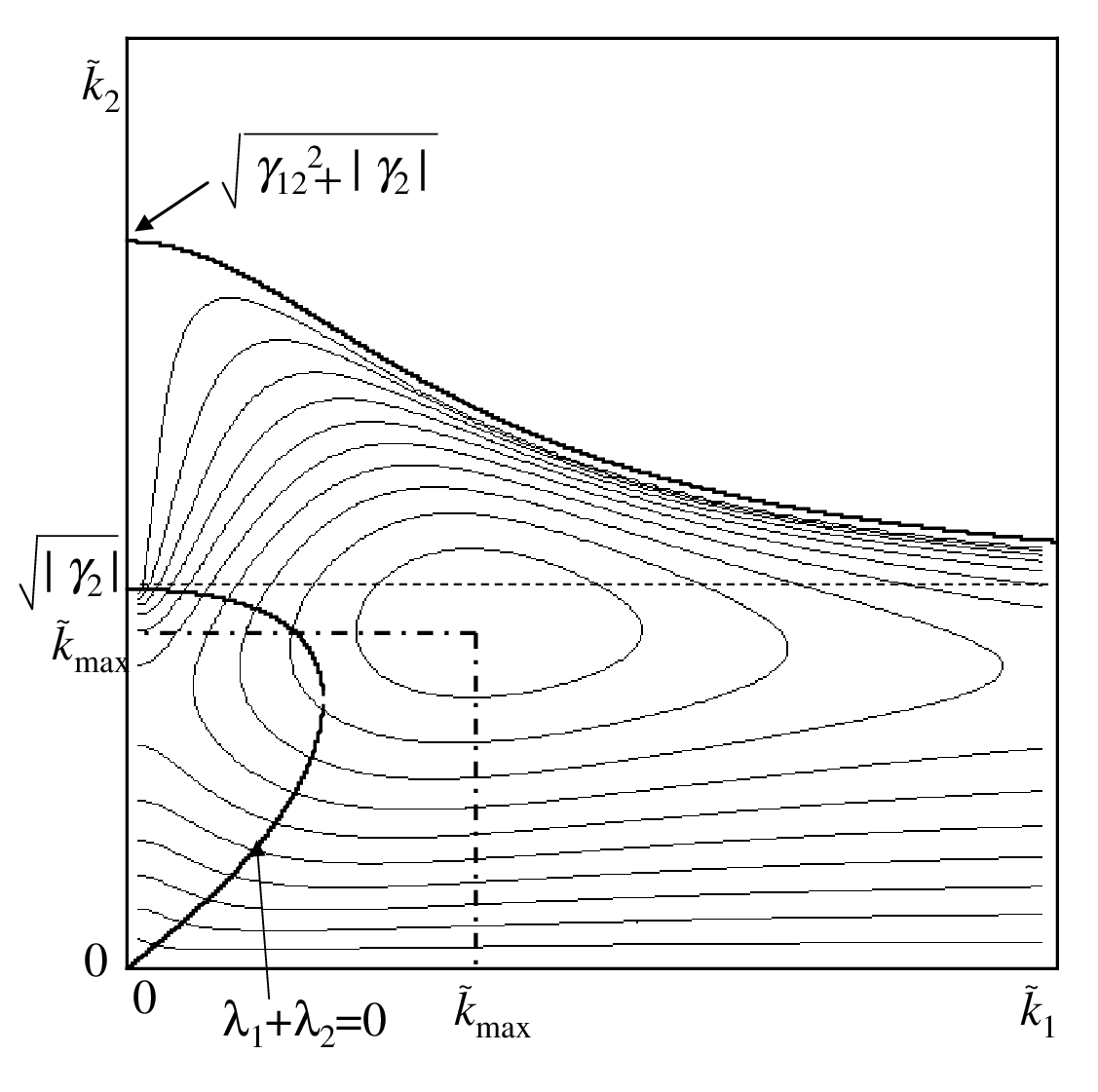}
\caption{The modulationally unstable region in $\tilde{k}_{1}$-$\tilde{k}_{2}$ space for cases (c) and (d) in Table \ref{coupclas}. The unstable region lies below the thick line given by $\tilde{k}_{2} = \sqrt{\gamma_{12}^{2} /(\tilde{k}_{1}^{2}+1) - \gamma_{2}} $. The boundary curve crossing the origin (bottom left) is given by $\Lambda_{1}+\Lambda_{2}=0$; the right (left) region from the boundary represents the region (i) ((ii)) (see the text). The contour plot shows the magnitude of the gain spectra $G={\rm Im} \Omega$ in arbitrary unit.}
\label{pararegb}
\end{figure}
Next, we consider the situation in which one component has an attractive intracomponent interaction. As shown in Sec. \ref{singleMIrev}, the $\psi_{2}$-component always undergoes the MI induced by the attractive force. It is interesting to consider how the presence of the other component affects the MI condition. The unstable region in $\tilde{k}_{1}$-$\tilde{k}_{2}$ space and the gain spectra $G={\rm Im} \Omega$ is shown in Fig. \ref{pararegb}. For $\gamma_{2}<0$ there appears to be a region satisfying $\Lambda<0$, where $\Omega_{+}^{2}$ can become negative (i.e., unstable) if $\Delta<0$. However, there is no region where these two inequalities ($\Lambda<0$, $\Delta<0$) are satisfied simultaneously. Hence, we will focus on the instability associated with $\Omega_{-}^{2}$. 

For $\gamma_{12}=0$, independent of the values of $\tilde{k}_{1}$, there is an ``unstable band" with $G=\frac{\hbar \tilde{k}_{2}}{2 m \ell^{2}} \sqrt{\gamma_{2} - k_{2}^{2}}$ in the range $0<\tilde{k}_{2}<\sqrt{|\gamma_{2}|}$. For $\gamma_{12} \neq 0$ the unstable region extends beyond the boundary line $\tilde{k}_{2}=\sqrt{|\gamma_{2}|}$. Two unstable regions exist: (i) $\Lambda>0$ with $\Delta > 0$ and (ii) $\Lambda<0$ with $\Delta > 0$. The condition for region (i) is the same condition as that for case (a) except the boundary line $\tilde{k}_{2} = \sqrt{\gamma_{12}^{2}/(k_{1}^{2}+1) + |\gamma_{2}| }$ has no intersection with the $\tilde{k}_{1}$-axis because $\gamma_{2}<0$. The condition for (ii) yields the imaginary $\Omega_{-}$ in the region bound by the curves $\tilde{k}_{1}=0$ and $\lambda_{1} + \lambda_{2}=0$ in Fig. \ref{pararegb}, which is comprised by the region given by (i). As a result, the MI condition is solely determined by $\Delta>0$ as in case (a).

In region (c), the system undergoes MI induced by both the intracomponent interaction $u_{2}$ and the intercomponent one $u_{12}$. The most unstable wave number and the maximum growth rate are again given by Eqs. (\ref{maxwavenum}) and (\ref{maxgrowth}). When $\gamma_{12}=0$, we reproduce the results of the single-component case $G_{\rm max}=\hbar |\gamma_{2}|/4m\ell = |u_{2}|n_{20}/\hbar$. Therefore, an increase of $\gamma_{12}$ always increases the unstable growth rate over that of the single-component case. When $G_{\rm max}$ is located within the unstable band $0<\tilde{k}_{2}<\sqrt{|\gamma_{2}|}$, the dominant contribution to the MI should be the intracomponent attraction associated with the negative $u_{2}$. If $\gamma_{12}>|\gamma_{2}| (2 |\gamma_{2}| + 1)$, the location of $G_{\rm max}$ moves outside the unstable band, so that the intercomponent repulsion $u_{12}$ has the dominant influence on the MI. Compared with the case (a), the magnitude of the gain spectra is larger by a few factor of about ${\cal O}(1)$. This is due to the multiplication effect of those two instabilities. 

Once the instability starts, the modulation develops by following the eigenvectors associated with the eigenvalue $\Omega_{-}$. According to Eq. (\ref{eigenvecmatrix}), the modulation becomes out-of-phase.

\subsubsection{Region (d): $\gamma_{2}<0$ and $\gamma_{12}<0$}
This region is similar to that of (c). The MI condition, the most unstable wave number and the corresponding maximum gain spectra are the same as those in region (c). From the eigenvectors of the modulation amplitudes, the MI is related to an in-phase evolution of the two-component modulation. 

\section{NUMERICAL SIMULATIONS}\label{numeri}
\subsection{Formulation of the simulations}
In this section, we present the results of our numerical simulations that illustrate the effect of the MI on the nonlinear evolution of the condensates. First, we describe the formulation, the initial conditions and the parameters of the simulations in detail. 

\subsubsection{The dimensionless GP equations with particle loss terms}
To reduce the number of the parameters, we assume $m_{1}=m_{2} \equiv m$ and $\omega_{1}=\omega_{2} \equiv \omega_{\perp}$, so that $b_{1}=b_{2} \equiv b_{\rm ho}=\sqrt{\hbar / m \omega_{\perp}}$. It is  also convenient to introduce the scales characterizing the trapping potential $\omega_{\perp}^{-1}$, $b_{\rm ho}$ and $\hbar \omega_{\perp}$ for time, length and energy, respectively. By replacing the wave function with the total particle number $N$ $(=N_{1}+N_{2})$ as $\psi_{i} \rightarrow \psi_{i} \sqrt{N /b_{\rm ho}}$, the one-dimensional GP equations (\ref{2tgpe1D}) reduce to 
\begin{subequations}
\begin{eqnarray}
i \frac{\partial \psi_{1}}{\partial t} = \biggl[ -\frac{1}{2} \frac{\partial^{2}}{\partial z^{2}} 
+\frac{\lambda^{2}}{2}z^{2} + U_{1} |\psi_{1}|^{2} + U_{12} |\psi_{2}|^{2} \biggr] \psi_{1}, \\
i \frac{\partial \psi_{2}}{\partial t} = \biggl[ -\frac{1}{2} \frac{\partial^{2}}{\partial z^{2}} 
+\frac{\lambda^{2}}{2}z^{2} + U_{2} |\psi_{2}|^{2} + U_{12} |\psi_{1}|^{2} \biggr] \psi_{2}, 
\end{eqnarray} \label{2tgpendim}
\end{subequations}
where $U_{i} = u_{i} N / \hbar \omega_{\perp} b_{\rm ho} = 2 N a_{i}/b_{\rm ho}$ ($i=1,2$), $U_{12} = u_{12} N / \hbar \omega_{\perp} b_{\rm ho} = 2 N a_{12}/b_{\rm ho}$, and the normalization of the wave function is $\int d z |\psi_{i}(z)|^{2} = N_{i}/N  $. 

We will simulate the dynamics for the condensate with an attractive interaction. Therefore, we should include an effect of the atomic loss due to inelastic collisions \cite{Kagan}. We model this effect by adding to Eqs. (\ref{2tgpendim}) the phenomenological loss term 
\begin{equation}
{\rm loss} \hspace{2mm} {\rm term} = -i  \left(  \tilde{L}_{j}^{(3)} |\psi_{j}|^{4} + \tilde{L}_{\rm dif} |\psi_{3-j}|^{2}  \right) \psi_{j} \hspace{5mm} j=1,2 
\label{lossphen}
\end{equation}
with
\begin{equation}
\tilde{L}_{j}^{(3)} = \frac{1}{3!} \frac{L_{j}^{(3)} N^{2}}{6 \pi^{2} \omega_{\perp} b_{\rm ho}^{6}},  \hspace{3mm}  
\tilde{L}_{\rm dif} = \frac{1}{2!}  \frac{L_{\rm dif} N}{4 \pi \omega_{\perp} b_{\rm ho}^{3}} 
\end{equation}
in the right-hand side of Eq. (\ref{2tgpendim}). The first term on the right side of Eq. (\ref{lossphen}) is related with the three-body inelastic collisions, which is the dominant mechanism of particle loss when the self-focusing collapse of the condensate occurs. The second term on the right side of Eq. (\ref{lossphen}) represents the inelastic loss due to collisions between different components, which is associated with inelastic collision between different atomic species \cite{Ferrari} or the spin exchange collision for the two components with different hyperfine states \cite{Esry}. Because the detail of the particle loss through the collapse are not needed here, we use a value for $L_{j}^{(3)}$ and $L_{\rm dif}$ that is consistent with experimental results: $L_{j}^{(3)} = 1 \times 10^{-26} $ cm$^{6}$/s \cite{Savage} and $L_{\rm dif} = 3 \times 10^{-14}$ cm$^{3}$/s \cite{Esry}. 

\subsubsection{The initial conditions}
We numerically solved the time-dependent GP equations (\ref{2tgpendim}) using a Crank-Nicolson implicit scheme with $8 \times 10^{3}$ grid points and a time grid $\Delta t = 5.0 \times 10^{-4}$. The focus here is on how the MI grows spontaneously from the miscible condensates. Therefore, the initial two components should be uniform and overlap as much as possible in the trapping potential. To do this, we first prepared the stationary solution of $\psi_{1}$ component (denoted by $\psi_{\rm ini}$) by propagating Eq. (\ref{2tgpendim}a) in imaginary time, under the normalization $\int dz |\psi_{1}|^{2} =1$. This was done without the $U_{12}$-term and the particle loss term. Next, at $t=0$ in real time simulations, some fractions of $\psi_{1}$ component were suddenly put into the $\psi_{2}$ component that had the same density profile as $\psi_{1}$. Initially, each component has the same inverted-parabola density profile, but a different normalization condition $\int dz |\psi_{i}|^{2} = N_{i}/N$. If $U_{1}=U_{2}=U_{12}$ is not satisfied, this initial configuration is nonstationary and thus can develop following the MI. Hence, we focus on such nonstationary cases. This situation can be realized experimentally by using a rf-pulse to transfer the population from one hyperfine level of atoms to the other one \cite{Hall,Miesner}. 

\subsubsection{Parameters and the validity of the one-dimensional simulations}
We consider the quasi-one-dimensional geometry characterized by the aspect ratio $\lambda=0.02$. Using the mass of rubidium atoms and the radial trapping frequency $\omega_{\perp} =2 \pi \times 100$ Hz, we obtain the length scale $b_{\rm ho}=1.1$ $\mu$m and the time scale $\omega_{\perp}^{-1}=1.59$ msec. According to the values of typical alkali atoms, we fix the intracomponent s-wave scattering lengths for the simulations of the cases (a) and (b) in Table \ref{coupclas} as
\begin{equation}
 a_{1}=5.5 \hspace{1mm} {\rm nm}, \hspace{5mm} a_{2}=5.8 \hspace{1mm} {\rm nm}
 \end{equation}
and for the cases (c) and (d) as 
\begin{equation}
a_{1}=5.5 \hspace{1mm} {\rm nm}, \hspace{5mm} a_{2}=-0.2 \hspace{1mm} {\rm nm}. 
\end{equation}
Further simplification can be obtained if we confine ourselves to distribute the equal particle number for two components $N_{1}=N_{2}$, i.e., $\int dz |\psi_{i}|^{2}=1/2$ ($i=1,2$). Thus, we have the total particle number $N$ and the value of the intercomponent s-wave scattering length $a_{12}$ as variable parameters.

We used the quasi-one-dimensional model under the assumption that the transeverse motion would be frozen. To justify this assumption, the energy scale of the transverse confinement should be much larger than the nonlinear interaction energy. This yields the condition $a_{i} N / b_{\rm ho} \ll 1$ \cite{Garcia3}. Unfortunately, this condition is not satisfied for our parameters; we obtain $a_{i} N / b_{\rm ho} \sim 1$ for the typical parameters presented below. However, the resulting transverse motion is only a rapid breathing oscillation, which does not affect the MI-induced dynamics in the longitudinal direction \cite{Kasamatsu}. The more critical condition for our study is to prevent the {\it transverse collapse} \cite{Carr}, being given by $|8 \pi a_{i} N |\psi_{i}(z)|^{2}/b_{\rm ho}| < 11.7$. In our situation, this position-dependent condition is nearly satisfied whenever the focusing collapse of the attractive condensates occurs at the trapping center in the following discussion. 

\subsection{Numerical results for region (a)}
We first consider region (a) in which all three scattering lengths are positive. The initial state is given by the miscible condensates with the same density profile of inverted parabolas. This region, together with the initial state, coincides exactly with the experiment of Miesner et al. \cite{Miesner}. They first prepared all $^{23}$Na atoms in the $|F=1, m_{F}=1 \rangle$ hyperfine state in an optical trap and then placed instantaneously half of them into the $|F=1, m_{F}=0 \rangle$ state using an rf field. Letting the system evolve freely while using the quadratic Zeeman effect to prevent the $|F=1, m_{F}=-1 \rangle$ component from appearing, they found that spin domains formed with two components alternatively aligned from the initially miscible condensates. Our previous paper \cite{Kasamatsu} pointed out that this observation is due to the MI caused by the intercomponent repulsive interaction. In the following, we describe the features of the nonlinear dynamics in more detail using one-dimensional simulations. 

\subsubsection{Dynamical features}
\begin{figure}
\includegraphics[height=0.41\textheight]{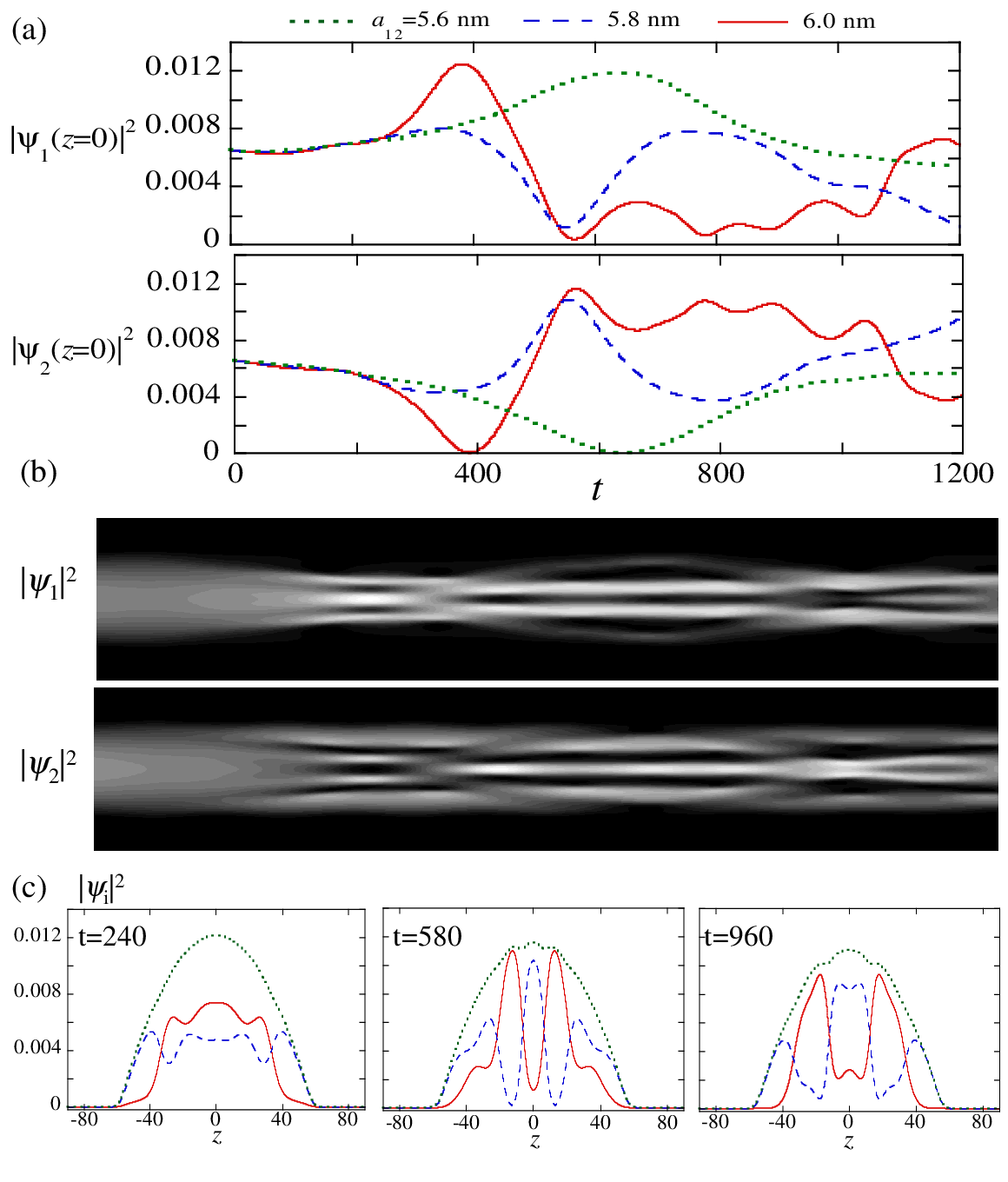}
\caption{(Color online.) Numerical results of Region (a) for $N=5 \times 10^{3}$. (a) Time development of condensate density $|\psi_{1}|^{2}$ (upper panel) and $|\psi_{2}|^{2}$ (lower panel) at $z=0$ for $a_{12}=5.6$ nm, 5.8 nm, and 6.0 nm. (b) The contour plots of the density of both components with respect to time ([0,1200], horizontal axis) and $z$ ([-100,100], vertical axis) for $a_{12}=6.0$ nm. (c) The density profiles of $|\psi_{1}|^{2}$ (solid-curve), $|\psi_{2}|^{2}$ (dashed-curve) and total density $n_{T}=|\psi_{1}|^{2}+|\psi_{2}|^{2}$ (dotted-curve) for $a_{12}=6.0$ nm at $t=240$, $t=580$, and $t=960$. The unit of time is $\omega_{\perp}^{-1}$. }
\label{domedome}
\end{figure}
The MI changes greatly the behavior of nonstationary development of the condensates. In Fig. \ref{domedome}, we show the results of the numerical simulation for $N=5 \times 10^{3}$. Figure \ref{domedome}(a) represents the development of the condensate density at the center ($z=0$) for several values of $a_{12}$. A crucial difference of the dynamical behavior is seen across the critical value about $a_{12}^{c}=\sqrt{a_{1} a_{2}} = 5.65$ nm, which corresponds to the criterion for phase separation. When $a_{12}$ is smaller than the critical value $a_{12}$, $|\psi_{1}(0,t)|^{2}$ ($|\psi_{2}(0,t)|^{2}$) first increases (decreases) gradually and makes a slow oscillation. Because $a_{2}>a_{1}$ and $a_{12}>0$, the density of the $\psi_{1}$ component is located at the center, surrounded by that of the extended $\psi_{2}$ component. As $a_{12}$ increases, the oscillation becomes nonperiodic; the amplitude of $|\psi_{1}|^{2}$ drops to zero after some time [solid and dashed curves in Fig. \ref{domedome}(a)], which represents that $\psi_{1}$ is replaced by $\psi_{2}$ at the center. This replacement shows the onset of the MI. 

The dynamical process of the spatial pattern formation induced by MI is clearly seen in the evolution of the overall density. Figure \ref{domedome} (b) and (c) show the evolution of each condensate density for $a_{12}=6.0$ nm exceeding the critical value. Throughout the dynamics, the modulation of the density causes two components to become out-of-phase. Hence, the total density $n_{T}=|\psi_{1}|^{2}+|\psi_{2}|^{2}$ keeps approximately its initial shape in spite of the irregular profile of each component, as shown in Fig. \ref{domedome}(c). After $t=300$ the density breaks up into smaller domains. The domains of the two components alternate in location, while the total density hardly changes even after the domain formation. Because $a_{1}<a_{2}$, the occupied region of $|\psi_{2}|^{2}$ expands rather than $|\psi_{1}|^{2}$ as seen in Fig. \ref{domedome}(b) and (c). The inelastic loss shrinks the size of both components monotonically. 

\begin{figure}
\includegraphics[height=0.41\textheight]{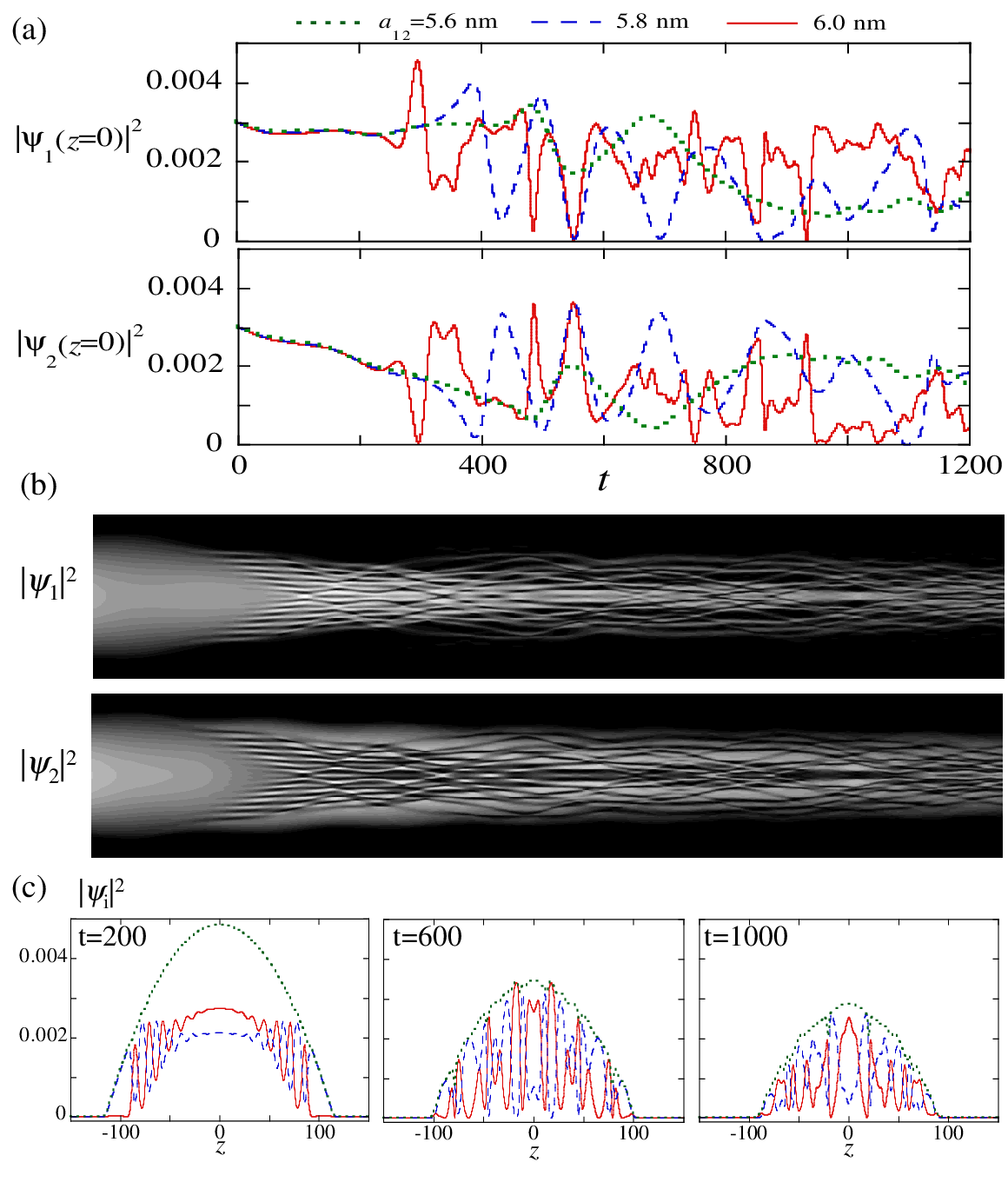}
\caption{(Color Online.) Numerical results of region (a) for $N = 5 \times 10^{4}$. (a) Time development of condensate density $|\psi_{1}|^{2}$ (upper panel) and $|\psi_{2}|^{2}$ (lower panel) at $z=0$ for $N=5 \times 10^{4}$ and $a_{12}=5.6$ nm, 5.8 nm, and 6.0 nm. (b) Contour plots of the density of both components with respect to time ([0,1200], horizontal axis) and $z$ ([-160,160], vertical axis) for $a_{12}=6.0$ nm. (c) The density profiles of $|\psi_{1}|^{2}$ (solid-curve), $|\psi_{2}|^{2}$ (dashed-curve) and total density $n_{T}=|\psi_{1}|^{2}+|\psi_{2}|^{2}$ (dotted-curve) for $a_{12}=6.0$ nm at $t=200$, $t=600$, and $t=1000$. The unit of time is $\omega_{\perp}^{-1}$. }
\label{domedome2}
\end{figure}
When the initial particle number increased, the dynamics become more dramatic. Figure \ref{domedome2}(a) shows the time evolution of the central density for $N=5 \times 10^{4}$. Compared with Fig. \ref{domedome}(a), the central density makes a more rapid and complex oscillation after the MI occurs. Then, the condensates form much more domains than those for $N=5 \times 10^{3}$. Figures \ref{domedome2}(b) and (c) show that the amplitude of the density modulations grows first near the edge of the condensate. This growth proceeds from the edge to the center, leading to the formation of localized condensate domains. Since the total number $N$ is large, the inelastic loss shrinks the condensate size much faster than the case of smaller $N$.


Spatially localized domains can be created even in condensates with a repulsive interaction. This is a salient feature in a multicomponent system; in the case of a scalar condensate (without a periodic potential), localized domains such as bright solitons are created only when the interaction nonlinearity is attractive. Actually, the generated domains have a solitary wave structure, where the spatial distribution of the condensate phase $\theta_{i}=$arg$\psi_{i}$ is almost flat within each domain and its value jumps across the density dips where the domains of the other exist \cite{Kasamatsu}. 

After multiple domains form, the dynamics of each domain may be determined by the phase-dependent interaction between {\it intracomponent} domains and the density-dependent interaction between {\it intercomponent} domains. It is known that the interaction between the bright solitons in a single component BEC depends on their phase difference \cite{Salasnich,Carr}. In our simulation, the evolution of the phase difference between domains is determined nontrivially, following the nonlinear dynamics caused by the MI. When two domains of the same component approach and share the same spatial location, the domains exchange particles. However, the domain of the other component blocks this approach because of the repulsive mean-field interaction between domains of different components. This is a phenomena analogous to the Josephson effect, predicted in Ref. \cite{Ao2}, where two single-component condensates with the different phases are separated by a potential barrier. The oscillations in Figs. \ref{domedome} and \ref{domedome2} after the MI occurred may be caused by the cooperative oscillation of the two-component soliton trains by this Josephson effect, but this needs further investigation.

\subsubsection{Comparison with the MI analysis}
The above dynamics were triggered by the MI induced by the intercomponent coupling $U_{12} \propto a_{12}$. The first modulation growth is determined by the fastest growth mode that has the most negative value of $\Omega_{-}^{2}$ of Eq. (\ref{freqkika}). The corresponding wave number $\tilde{k}_{\rm max}$ and the maximum gain $G_{\rm max}$ will determine the early behavior of the dynamics such as the modulation growth time and the number of initially created domains. If we assume the homogeneous condensates, they are given by Eqs. (\ref{maxwavenum}) and (\ref{maxgrowth}); in the units of this section, they are 
\begin{eqnarray}
\tilde{k}_{\rm max} = \left[ U_{1} n_{10} \left( \sqrt{(\gamma_{2}-1)^{2}+4 \gamma_{12}^{2}}-\gamma_{2}-1 \right)  \right]^{1/2} \label{kmaxdimless} \\
\frac{G_{\rm max}}{\omega_{\perp}} = 2 U_{1}^{2} n_{10}^{2}  \left( \sqrt{(\gamma_{2}-1)^{2}+4 \gamma_{12}^{2}}-\gamma_{2}-1 \right) \label{gmaxdimless}
\end{eqnarray}
with $\gamma_{2}=a_{2} n_{20}/a_{1} n_{10}$ and $\gamma_{12}=(a_{12}/a_{1}) \sqrt{n_{20}/n_{10}}$. To estimate $\tilde{k}_{\rm max}$ and $G_{\rm max}$, we assume that the density profile of the initial $\psi_{1}$ component has the one-dimensional Thomas-Fermi profile $n_{\rm ini}= |\psi_{\rm ini}|^{2} =( \mu_{1} - \lambda^{2} z^{2} / 2)/U_{1}$ with $\mu_{1}=(3 \lambda a_{1} N/ 2 \sqrt{2} b_{\rm ho})^{2/3}$. Because half of the $\psi_{1}$ component is suddenly transferred to $\psi_{2}$, we use the density $n_{\rm ini}/2$ at $z=0$ as an approximation of $n_{i0}$ ($i=1,2$) in Eqs. (\ref{kmaxdimless}) and (\ref{gmaxdimless}). For example, the parameters in Fig. \ref{domedome}(b) and (c) yield $\tilde{k}_{\rm max} = 0.205$, $G_{\rm max}/\omega_{\perp} = 0.0275$, and those in Fig. \ref{domedome2}(b) and (c) yield $\tilde{k}_{\rm max} = 0.441$, $G_{\rm max}/\omega_{\perp} = 0.592$. The quantity $2R_{z}/(2\pi/k_{\rm max})$ is approximately the number of generated domains in the simulation. This quantity equals 3.7 for $N=5 \times 10^{3}$ and 17.3 for $N=5 \times 10^{4}$, in reasonable agreement with the numerical results. The growth time is determined as $2\pi/G_{\rm max}$, which gives 229 (in units of $\omega_{\perp}^{-1}$) for $N=5 \times 10^{3}$ and 10.6 for $N=5 \times 10^{4}$. These times approximate the time scale for the first rapid growth of the central density shown in Fig. \ref{domedome}(a) and Fig. \ref{domedome2}(a). At later times, the linear analysis is not applicable. 

\subsubsection{Analogy to the dynamics of condensates with attractive interactions}
We point out that the dynamics described here is analogous to the collapse dynamics and soliton-train formation in a BEC with attractive interactions \cite{Salasnich,Kamchatnov,Carr}. This analogy is as follows. The total density $n_{T}=|\psi_{1}|^{2}+|\psi_{2}|^{2}$ hardly changes during the time evolution as seen in Figs. \ref{domedome}(c) and \ref{domedome2}(c). Thus, we can rewrite the dynamical equations (\ref{2tgpendim}) as 
\begin{subequations}
\begin{eqnarray}
i \frac{\partial \psi_{1}}{\partial t } = \biggl( -\frac{1}{2} \frac{\partial^{2}}{\partial z^{2}} + \frac{\lambda^{2}}{2} z^{2} + \frac{1}{2} (U_{1}+U_{12}) n_{T} \nonumber \\
+ \frac{1}{2} (U_{1}-U_{12}) |\psi_{1}|^{2} - \frac{1}{2} (U_{1}-U_{12}) |\psi_{2}|^{2}  \biggr) \psi_{1}, \\
i \frac{\partial \psi_{2}}{\partial t } = \biggl( -\frac{1}{2} \frac{\partial^{2}}{\partial z^{2}} + \frac{\lambda^{2}}{2} z^{2} + \frac{1}{2} (U_{2}+U_{12}) n_{T} \nonumber \\
- \frac{1}{2} (U_{2}-U_{12}) |\psi_{1}|^{2} + \frac{1}{2} (U_{2}-U_{12}) |\psi_{2}|^{2}  \biggr) \psi_{2}.
\end{eqnarray}\label{analogeq}
\end{subequations}
In this formulation, the term $\lambda^{2} z^{2}/2 + (U_{i}+U_{12}) n_{T} / 2 \equiv V^{\rm eff}_{i}$ $(i=1,2)$ functions as the nearly static confining potential. Then, it is easy to understand how the nonlinear terms in Eq. (\ref{analogeq}) work. If $U_{1}<U_{12}$ in Eq. (\ref{analogeq}a), the intracomponent coupling becomes attractive, whereas the intercomponent coupling becomes repulsive. This favors a spatially localized structure of the $\psi_{1}$ component and phase separation between the two components. The same argument applies to the $\psi_{2}$ component. Even when $U_{2}>U_{12}$, the $\psi_{2}$ component forms a domain structure if $U_{1}<U_{12}$ is satisfied because the modulation develops out-of-phase.

This interpretation based on the condensates with attractive interactions can be extended to the effective one-component description of the domain formation \cite{Kasamatsu,Dutten}. Particularly, in the case of Stenger {\it et al.}'s experiments \cite{Stenger}, the two-component condensates of $^{23}$Na atoms are characterized by $U_{1}=U_{12} \equiv U$, in which Eq. (\ref{analogeq}) can be reduced to
\begin{subequations}
\begin{eqnarray}
i \frac{\partial \psi_{1}}{\partial t } = \biggl( -\frac{1}{2} \frac{\partial^{2}}{\partial z^{2}} + V_{i}^{\rm eff} \biggr) \psi_{1}, \\
i \frac{\partial \psi_{2}}{\partial t } = \biggl( -\frac{1}{2} \frac{\partial^{2}}{\partial z^{2}} + V_{i}^{\rm eff}  + (U_{2}-U) |\psi_{2}|^{2}  \biggr) \psi_{2}.
\end{eqnarray}
\label{effectiveonecom}
\end{subequations}
Then, Eq. (\ref{effectiveonecom}a) is a linear Schr\"{o}dinger equation and an effective attractive interaction appears for the $\psi_{2}$ component because $U_{2}-U<0$ \cite{Goldstein,Mueller}. This means that the sudden population transfer from $|1\rangle$ to $|2\rangle$ is formally equivalent to the sudden change of the atomic interaction of $| 2 \rangle$ from positive to negative. Therefore, the generated domains in Fig. \ref{domedome} and \ref{domedome2} may have a solitary wave structure such as a bright soliton train in a single component condensate \cite{Salasnich,Kamchatnov,Carr}. 
 
Some dynamics are also similar to those in the numerical simulation of bright soliton formation, in which bright solitons are first generated at the edge of an initial condensate \cite{Salasnich,Carr,Kamchatnov}. Since there is no noise in our simulation, the MI is triggered by self-interference fringes of the wave function. It was shown that the wavelength (amplitude) of self-interference fringes in the initial wave function is longer (larger) at the edge of the condensate than that at the central part \cite{Carr,Kamchatnov}. These fringes can be the seed of the modulation, first reaching the unstable wavelength $2 \pi/ \tilde{k}_{\rm max}$ at the edge. 
 
\subsection{Numerical results for region (b)}\label{regionbdyn}
We turn to the dynamics for the combination (b) in Table \ref{coupclas}. This combination of coupling constants does not appear in other systems described by similar model equations. The remarkable feature in this case is the existence of bright solitons supported by the {\it intercomponent attraction} even if the intracomponent interaction is {\it repulsive}. Some features such as stability and collisional properties of this new soliton were studied recently \cite{Garcias}. These studies also discussed the dynamics of soliton formation from the initial state that causes phase separation.

As in case (a), after preparing the initial states $\psi_{1}$ and $\psi_{2}$ that has an equivalent distribution with an inverted parabola and the normalization condition $\int dz |\psi_{i}|^{2}=1/2$ ($i=1,2$), we change instantaneously $a_{12}$ from zero to a negative value. In these simulations, small wave fragments with  a large kinetic energy are generated when the wave functions undergo self-focusing collapse. These waves spread to the edges of the area of numerical simulations and the reflected waves from the edges make the calculation unreliable. To prevent this reflection, we added an absorptive potential with the form $V_{\rm I}=V_{0} (1+\tanh[(z-z_{0})/\xi] \tanh[(z+z_{0})/\xi])$, where $z_{0}$ represents the position of the numerical edge and we set $V_{0} = 100 i$ and $\xi=5$; $V_{\rm I}$ can absorb only the waves that reach the edge. 

In this case, the dynamical evolution should be similar to what is observed in an attractive single-component BEC. This is due to the fact that the intercomponent attraction favors the spatial overlap of the wave function such that $|\psi_{1}|^{2} \simeq |\psi_{2}|^{2}$. Then, the coupled GP equations are reduced to 
\begin{equation}
i \frac{\partial \psi_{i}}{\partial t} \simeq \left( -\frac{1}{2} \frac{\partial^{2}}{\partial z^{2}} + \frac{\lambda^{2} z^{2}}{2} + (U_{i}+U_{12}) |\psi_{i}|^{2} \right) \psi_{i}  \hspace{3mm} i=1,2.
\end{equation}
Hence, we can expect that the MI occurs for $U_{i} + U_{12} < 0$ ($a_{i}+a_{12}<0$). Using the results in Fig. \ref{pararega}, the necessary condition for the MI is given by $a_{12} < a_{12}^{c} = -\sqrt{a_{1} a_{2}} =- 5.65$ nm. 

\subsubsection{Condensate dynamics in a harmonic potential}
\begin{figure}
\includegraphics[height=0.35\textheight]{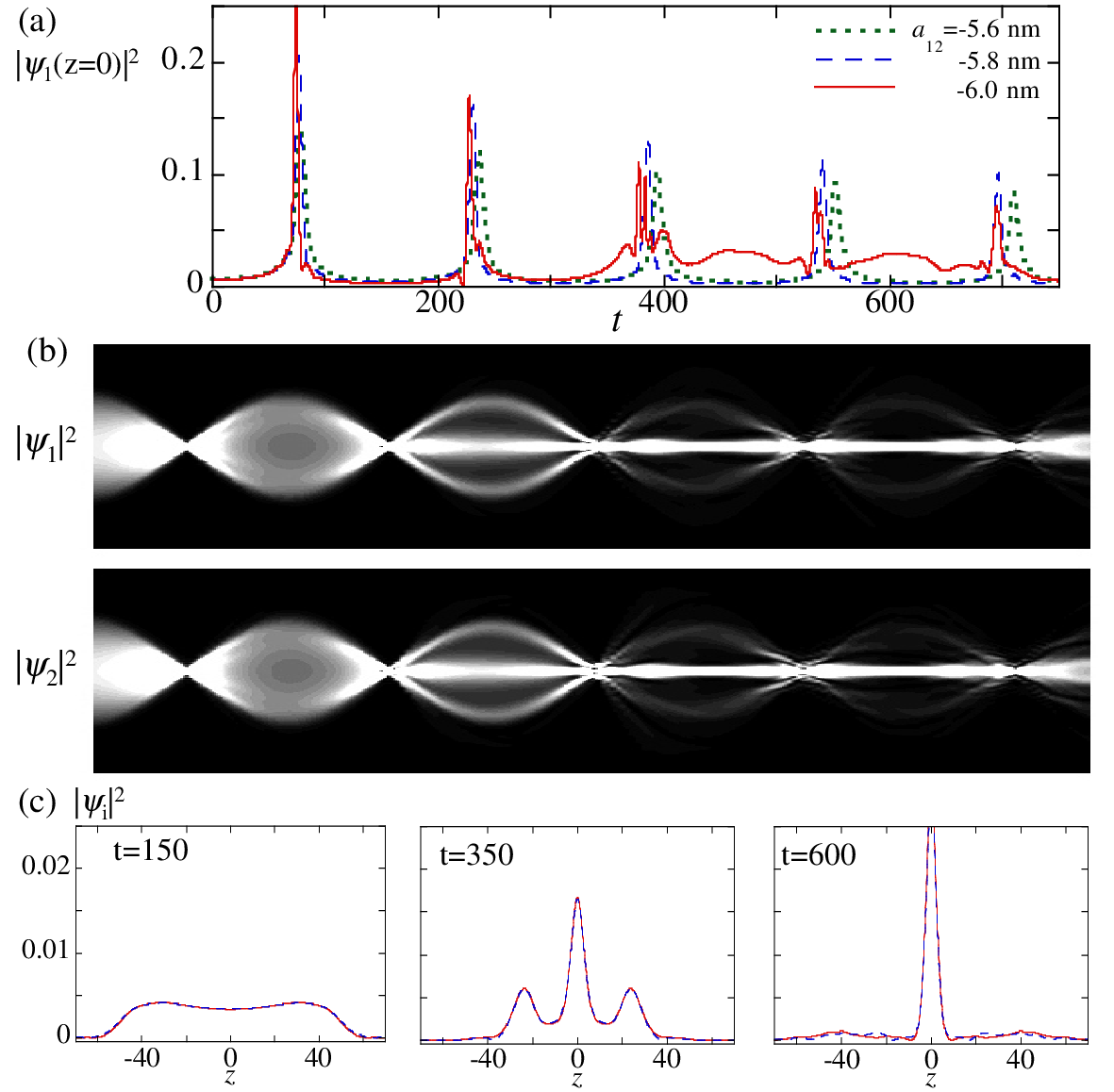}
\caption{(Color online.) Numerical results of region (b) for $N = 5 \times 10^{3}$, where the condensates are trapped by a harmonic potential. (a) Time development of condensate density $|\psi_{1}|^{2}$ at $z=0$ for $a_{12}=-5.6$, $-5.8$, and $-6.0$ nm. The insets show the details of the evolution during the first and third focusing. The development of $\psi_{2}$ component is similar to what is seen in (a) because the modulation develops in-phase. (b) The contour plots of the density of both components with respect to time ([0, 750], horizontal axis) and $z$ ([-100, 100], vertical axis) for $a_{12}=-6.0$ nm. (c) The density profiles of $|\psi_{1}|^{2}$ (solid-curve) and $|\psi_{2}|^{2}$ (dashed-curve) for $a_{12}=-6.0$ nm at $t=150$, $t=375$, and $t=600$. The unit of time is $\omega_{\perp}^{-1}$. }
\label{domedome3}
\end{figure}
The main feature of the attractive intercomponent interaction is to make the condensate self-focus on the center of the harmonic trap. Figure \ref{domedome3}(a) shows the time evolution of the central density $|\psi_{1}(0,t)|^{2}$ for $N=5 \times 10^{3}$ and several values of $a_{12}$. Due to the mutual attraction and the presence of the harmonic trap, the overall density contracts at the center. Subsequently, the kinetic-energy cost of this focusing makes the condensates expand with two components repeating this contraction and expansion. Above $a_{12} \simeq -5.7$ nm, although the central density of the condensates undergoes a large amplitude oscillation, no spatial fragmentation occurs. Below $a_{12} \simeq -5.7$ nm the central density collapses into some pulsed wave packets, or bright vector solitons, characterized by a sech-type form of both components \cite{Garcias}. The existence of these solitons are ensured by the intercomponent attractive interaction, because the intracomponent interaction is repulsive. The second self-focusing at $t \approx 220$ generates three solitary waves [see Fig. \ref{domedome3}(b)]. The soliton at the center does not move whereas the other small two solitons propagate outward and come back to the center because of the trapping potential. Then, the two propagating solitons merge with the central soliton and this merging generates a few bright solitons again. This nearly recurrent process repeats several times, creating  a single soliton at the center with the help of the inelastic particle loss.


\subsubsection{Condensate dynamics in an expulsive potential}
\begin{figure}
\includegraphics[height=0.25\textheight]{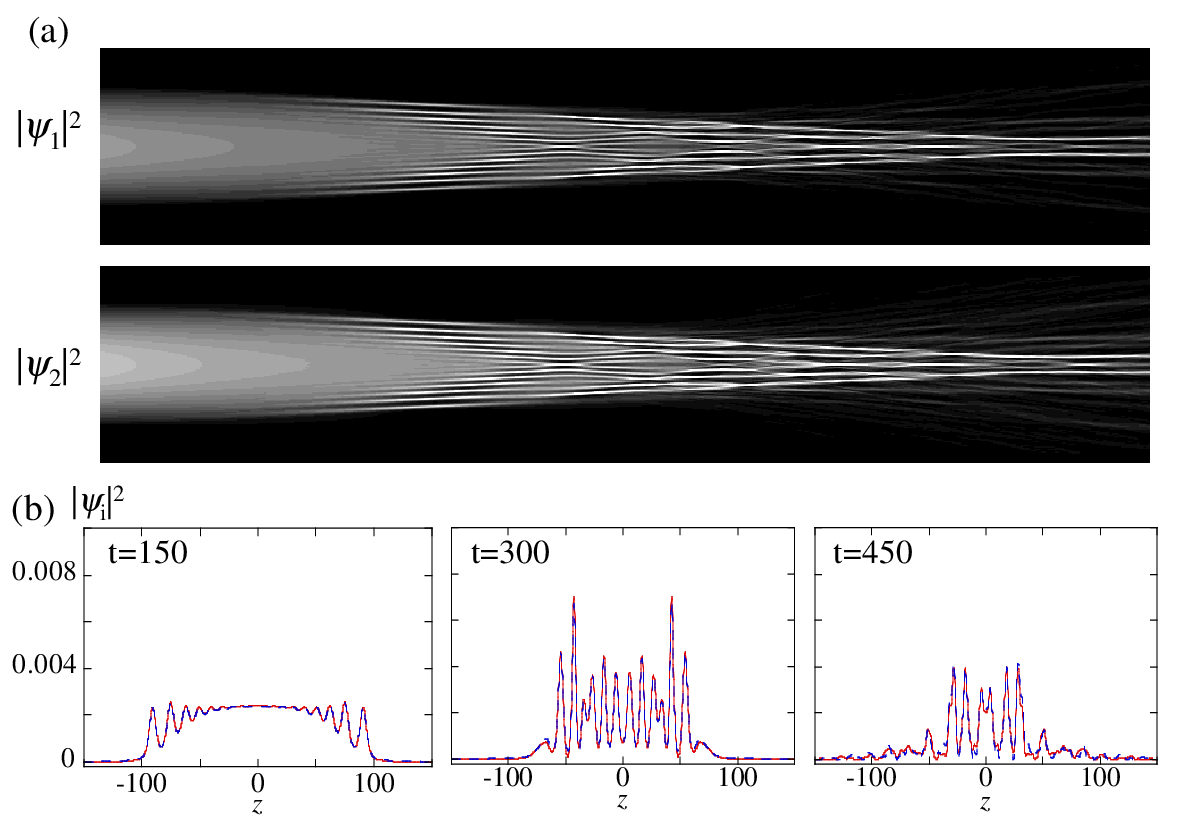}
\caption{(Color online.) Condensate dynamics in an expulsive potential. (a) The contour plots of the density of both components with respect to time ([0, 600], horizontal axis) and $z$ ([-200, 200], vertical axis) for $N=5 \times 10^{4}$ and $a_{12}=-6.0$ nm. (c) The density profiles of $|\psi_{1}|^{2}$ (solid-curve) and $|\psi_{2}|^{2}$ (dashed-curve) at $t=150$, $t=300$ and $t=450$. The unit of time is $\omega_{\perp}^{-1}$. }
\label{domedome35}
\end{figure}

It is not easy to compare the above numerical results with the MI analysis in Sec. \ref{moduinst}. The density inhomogeneity in the numerical simulation has a significant effect on the soliton formation, the MI analysis for the homogeneous condensate cannot be applicable. Also, the large particle loss of the first collapse makes the use of the particle number $N$ nontrivial to estimate $\tilde{k}_{\rm max}$. To prevent the focusing collapse, we ran a similar simulation but with a trapping potential with negative curvature, also known as an expulsive potential \cite{Khaykovich,Carr}. Figure \ref{domedome35} shows the resulting dynamics for a weak expulsive potential $V=-(0.1 \lambda)^{2} z^{2}/2$, where we used the parameters $N = 5 \times 10^{4}$ and $a_{12} = - 6.0$. In this case, a self-focusing collapse does not occur, but the density modulation grows spontaneously from the edges of the condensate, forming a bright soliton train. This result agrees with the single-component result in Ref. \cite{Carr}.

The MI condition in this case is given by the same analytic form in region (a). For example, the fastest growth mode is given by Eq. (\ref{kmaxdimless}) and the characteristic length scale $2\pi/\tilde{k}_{\rm max} = 14.2$ for $N=5 \times 10^{4}$ and $a_{12}=-6.0$ nm is in reasonable agreement with the wave length of the growing modulation and the size of the bright solitons [see Fig. \ref{domedome35}(b)] . 

\subsection{Numerical results for region (c)}
We now address the situation in which the intracomponent interaction in one component is repulsive, while that of the other is attractive. A single-component condensate with attractive interactions can generate bright solitary waves because of its nonlinear self-focusing effect \cite{Salasnich,Kamchatnov,Carr}. Here we ask the related question: in the presence of two components, how are the self-focusing collapse and the formation process of bright solitons affected by the {\it intercomponent} interaction?

The conditions for the numerical simulation are those for region (b). To prevent the reflection of the waves at the numerical edge, we used the absorbing potential $V_{\rm I}$. The initial states $\psi_{1}$ and $\psi_{2}$ were distributed equally with $N_{1}=N_{2}=N/2$ (i.e., $\int dz |\psi_{i}|^{2} = 1/2$ $(i=1,2)$). We changed the value of $a_{2}$ to a negative one $a_{2} = - 0.2$ nm at $t=0$ . Then, the $\psi_{2}$ component generates bright solitons via MI induced by the intracomponent attraction. If $a_{12}=0$, the time evolution of $\psi_{2}$ is the same with the single component problem. However, the presence of the second component and the resulting intercomponent interaction changes the dynamics significantly. In this subsection, we consider the case in which the intercomponent interaction is repulsive. 

\subsubsection{Condensate dynamics in a harmonic potential}
\begin{figure}
\includegraphics[height=0.41\textheight]{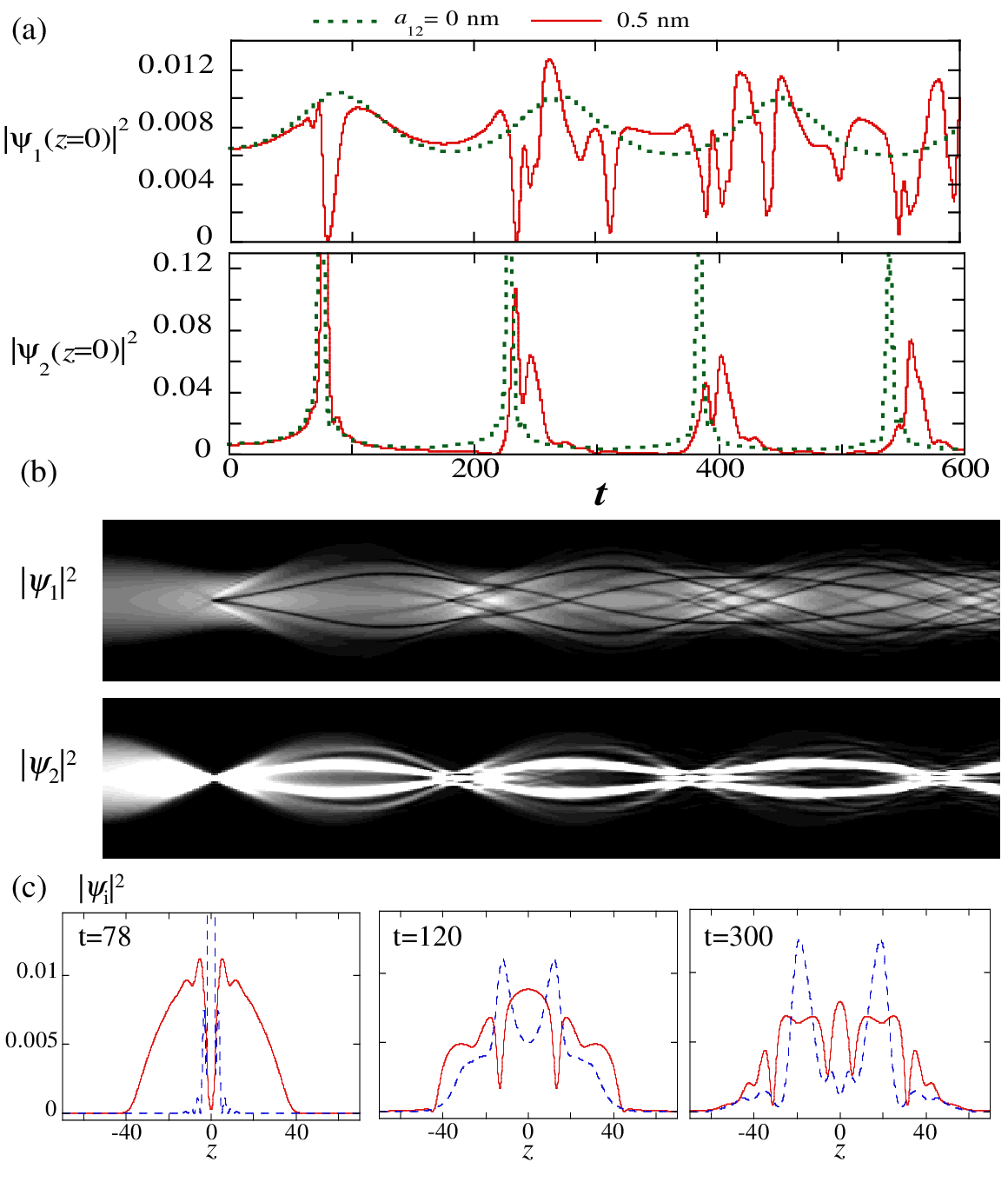}
\caption{(Color online.) Numerical results of region (c) for $N = 5 \times 10^{3}$, where the condensates are trapped by a harmonic potential. (a) Time development of condensate density $|\psi_{1}|^{2}$ (upper panel) and $|\psi_{2}|^{2}$ (lower panel) at $z=0$ for $a_{12}=0$ nm and 0.5 nm. (b) The contour plots of the density of both components with respect to time ([0, 600], horizontal axis) and $z$ ([-100, 100], vertical axis) for $a_{12}=0.5$ nm. (c) The density profiles of $|\psi_{1}|^{2}$ (solid-curve), $|\psi_{2}|^{2}$ (dashed-curve) for $a_{12}=0.5$ nm at $t=78$, $t=120$, and $t=300$. The unit of time is $\omega_{\perp}^{-1}$. }
\label{domedome4}
\end{figure}
We first ran the simulation in the presence of a harmonic potential for the particle number $N=5 \times 10^{3}$. The time evolution of the central density is shown in Fig. \ref{domedome4}(a). For $a_{12}=0$, while the repulsive $\psi_{1}$ component makes a breathing oscillation caused by a sudden population change at $t=0$, the $\psi_{2}$ component undergoes contraction and expansion as we found previously (Fig. \ref{domedome3}). However, the $\psi_{2}$ component forms no spatial pattern, probably because the particle number is not large enough to cause the instability. However, $a_{12}$ has a larger influence on the MI as it increases, eventually leading to the spatial pattern shown in Fig. \ref{domedome4}(b) and (c). This result shows the increase of the MI strength through the presence of the other component, as found in Sec. \ref{moduinst}. 

The numerical simulation reveals how the spatial pattern forms from the increased MI. Initially, the attractive $\psi_{2}$ component focuses at the center. Because of the intercomponent repulsion, the density modulation grows out-of-phase between the two components and the focused $\psi_{2}$ creates a density dip in $\psi_{1}$ at the center. This strong density disturbance generates counter-propagating density-kinks, also known as dark solitons, in the $\psi_{1}$ component. A similar formation mechanism for the single-component system was found in Ref. \cite{Burger2}, where the disturbance was given by an external potential. The rigidity of the dark solitons is ensured by the fact that they are clearly visible and propagate stably in the subsequent time evolution, as seen in Fig. \ref{domedome4}(b). The bright solitons of the attractive $\psi_{2}$ component also co-propagate, being embedded by these dark solitons. This composite soliton is referred to as ``dark-bright soliton" or ``gray-bright soliton", which is characteristic of the system having a vector order parameter \cite{Busch,Kevrekidis}. Though the composite solitons propagate outward, the bright solitons slip out of the density dips of the dark solitons, coming back first to the center. This causes the collision of the multiple bright solitons, which generates again the new composite solitons. A further increase in $N$ or $a_{12}$ increases the number of generating solitons and their collision gives rise to more fine-density ripples.

\subsubsection{Condensate dynamics in different trapping potentials}
Because the focus here is on nonlinear dynamics induced by the MI from initially miscible condensates, it is desirable that the two components are overlapped as much as possible while the MI occurs. To prevent the focusing collapse of the attractive component, we can also consider an expulsive potential. However, the use of the same expulsive potential for both components has a negative influence on the repulsive component because this component quickly expands and disappears. To avoid this problem, we use different trapping potentials $V_{1}=\lambda_{1}^{2} z^{2}/2$ and $V_{2}=\lambda_{2}^{2} z^{2}/2$ for the two components. This situation can be realized experimentally using the difference in the magnetic $g$-factor or the index of hyperfine sublevels for atoms in each component \cite{Ho}. Also, tuning the wavelength of an optical laser beam can create an optical potential that depends on the atomic hyperfine spin state \cite{Mandel}. 

\begin{figure}
\includegraphics[height=0.27\textheight]{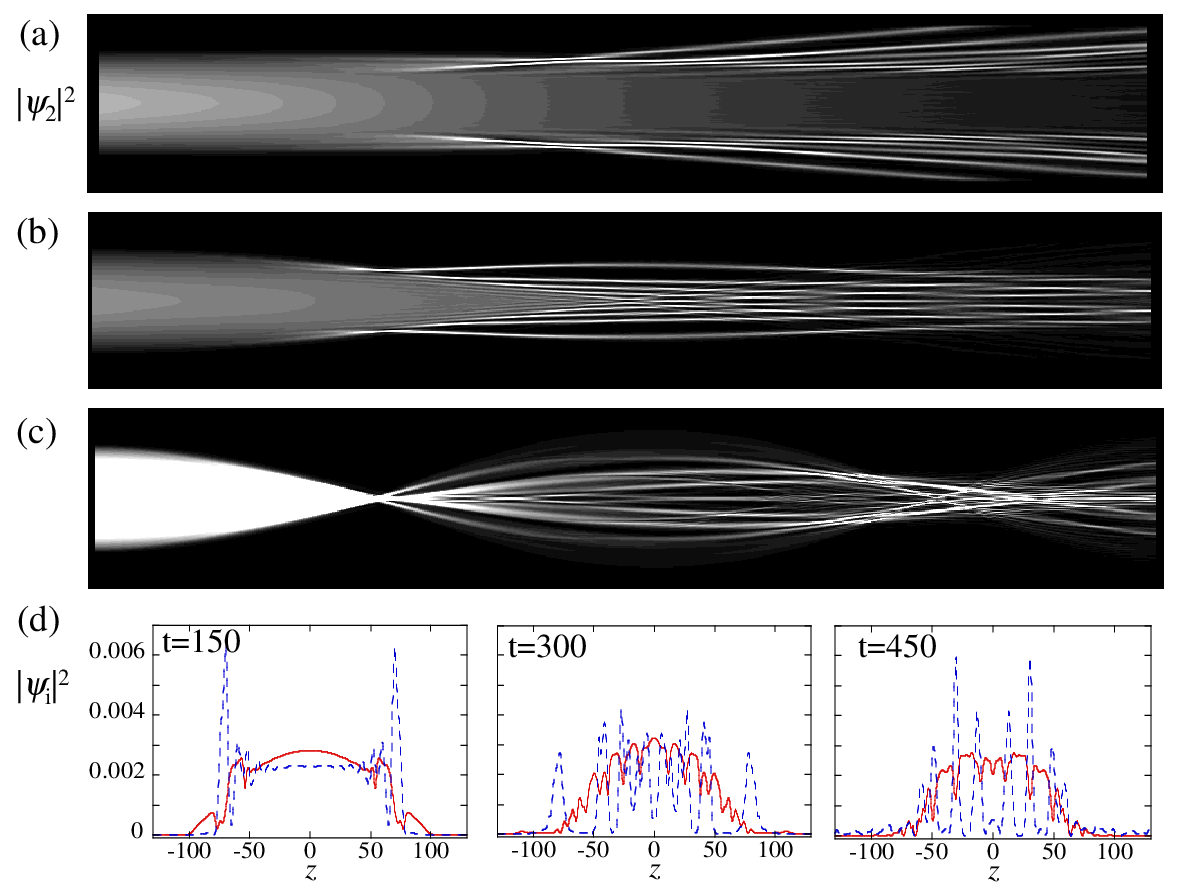}
\caption{(Color online.) Contour plots of the density of the attractive $\psi_{2}$ component with respect to time ([0, 500], horizontal axis) and $z$ ([-200, 200], vertical axis) for $N=5 \times 10^{4}$, $a_{12}=0.5$ nm, and (a) $\lambda_{2}=0$, (b) $\lambda_{2}=0.3 \lambda$, and (c) $\lambda_{2}=0.6 \lambda$ ($\lambda=0.02$). (d) The density profiles of $|\psi_{1}|^{2}$ (solid-curve), $|\psi_{2}|^{2}$ (dashed-curve) for the plot (b) at $t=150$, $t=300$, and $t=450$. The unit of time is $\omega_{\perp}^{-1}$. }
\label{domedome55}
\end{figure}
To better understand the role of the $\psi_{2}$ trapping frequency on the dynamics, we ran a simulation with the parameters $N=5 \times 10^{4}$ and $a_{12}=0.5$ nm and the same trapping frequency $\lambda_{1}=\lambda$ for $\psi_{1}$, but varied the trapping frequency for $\psi_{2}$. The results are shown in Fig. \ref{domedome55}. In (a), the time development of only the attractive $\psi_{2}$ component is shown for the case without a trapping potential $\lambda_{2}=0$. Because of the repulsive intercomponent interaction the $\psi_{2}$ component is pushed aside by the $\psi_{1}$ component. In addition, although the modulation grows from the edge, the bright solitons cannot move inside so they instead go outside. As a result, to cancel the effect of the intercomponent repulsion and to make the two components overlap, we had to use a trap with a weakly positive curvature for $\psi_{2}$. We found that for $\lambda_{2}=0.3\lambda$, the trapping potential $\lambda_{2}^{2} z^{2}/2$ and the intercomponent repulsion $U_{12} |\psi_{1}|^{2} \simeq U_{12} n_{\rm ini}/2$ are balanced, creating a flat effective potential. Then, focusing of $\psi_{2}$ at the center does not occur and we can thus study the pattern forming dynamics. Further increase in $\lambda_{2}$ focuses the $\psi_{2}$ component into the center as shown in Fig. \ref{domedome55}(c). 

From Fig. \ref{domedome55}(d), we find that the generated solitons also have the gray-bright character, where the bright solitons of $\psi_{2}$ combine with the density dips of $\psi_{1}$. The formation dynamics is similar to those found in the previous sections, in which the out-of-phase modulation grows from the condensate edges and evolves into the solitary waves. We find that the MI-induced dynamics is very sensitive to the change of $a_{12}$. For $a_{12} \neq 0$ both the characteristic time scale for the MI to start and the wave length of the initially developed modulation from the condensate edge [the left panel of Fig \ref{domedome55}(d)] decrease from those found for the simulation of $a_{12}=0$ (not shown). As a result, the size of the bright solitons become smaller with increasing $a_{12}$. This is the multiplication effect of the MI caused by the intercomponent interaction. 

\subsection{Numerical results for region (d)}
Finally, we discuss the dynamics when both $a_{2}$ and $a_{12}$ are negative. The numerical procedure is the same as that in the last section except this case has different scattering lengths. At $t=0$ we give $a_{2}=-0.2$ nm and some negative value of $a_{12}$. 

\subsubsection{Condensate dynamics in a harmonic potential} \label{aridesu}
\begin{figure}
\includegraphics[height=0.41\textheight]{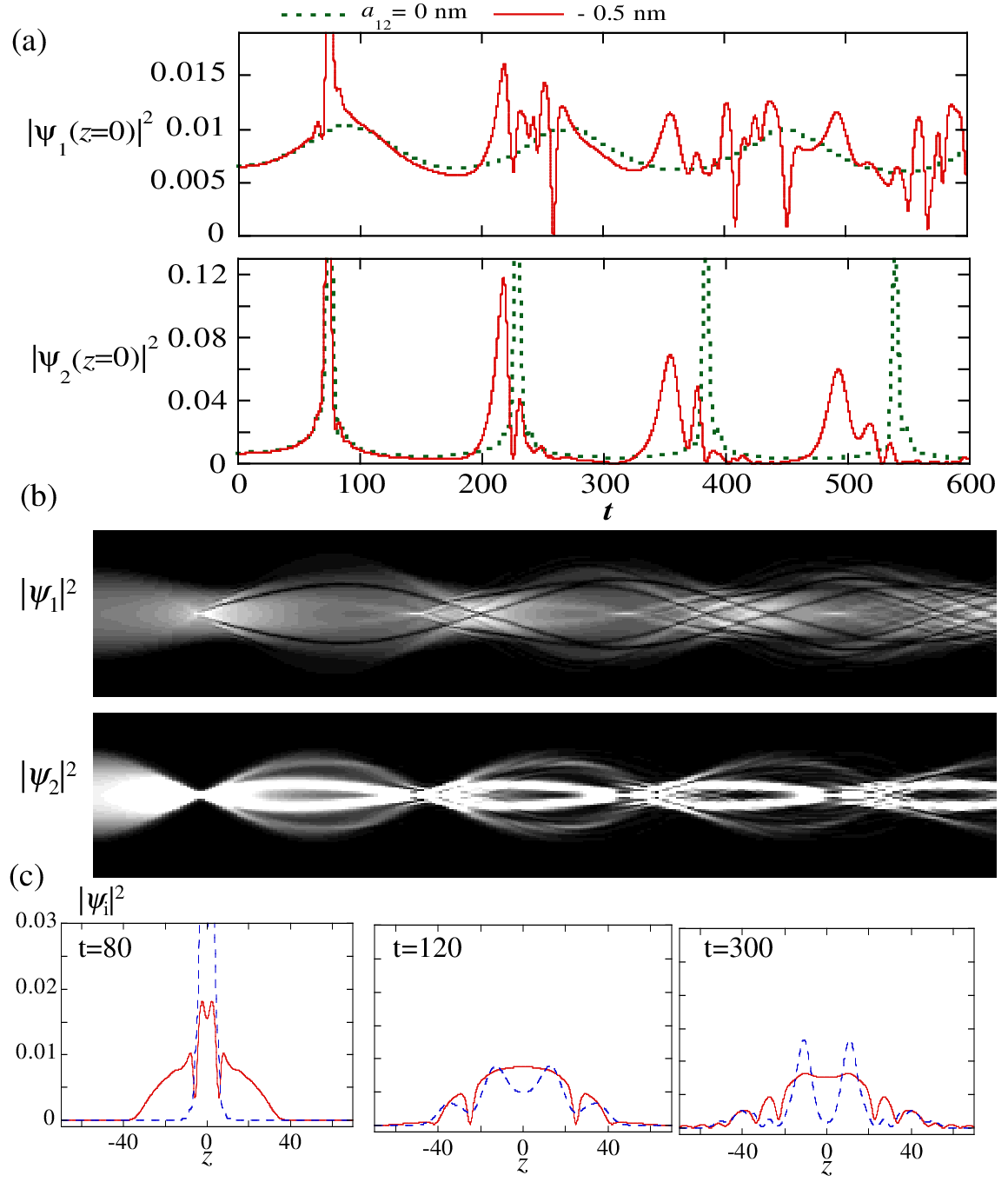}
\caption{(Color online.) Numerical results of region (d) for $N = 5 \times 10^{3}$, where the condensates are trapped by a harmonic potential. (a) Time development of condensate density $|\psi_{1}|^{2}$ (upper panel) and $|\psi_{2}|^{2}$ (lower panel) at $z=0$ for $N=5 \times 10^{3}$ and $a_{12}=0$ nm and $-0.5$ nm. (b) Contour plots of the density of both components with respect to time ([0, 600], horizontal axis) and $z$ ([-100, 100], vertical axis) for $a_{12}=-0.5$ nm. (c) The density profiles of $|\psi_{1}|^{2}$ (solid-curve) and $|\psi_{2}|^{2}$ (dashed-curve) for $a_{12}=-0.5$ nm at $t=80$, $t=120$, and $t=300$. The unit of time is $\omega_{\perp}^{-1}$. }
\label{domedome6}
\end{figure}
According to the MI analysis, the modulation should develop in-phase in this region. Thus, in the harmonic potential, the density focusing due to the attractive component creates a local density hump of the repulsive component. Both sides of this density hump evolve to a counter-propagating dark soliton pair in the repulsive component, as seen in Fig. \ref{domedome6}(b). Before these dark solitons come back to the center, the second focusing collapse occurs and generates a new dark soliton pair. On the other hand, the focusing collapse also generates counter-propagating bright solitons in the attractive component. This combined dynamical process and the resulting multiple collisions of the solitons at the center make the dynamics extremely complex. As the particle number is increased, although the above dynamical feature seems to be similar, we cannot obtain a clear physical picture and the reliable numerical accuracy because of the rapid density fluctuations generated through the above processes.

As seen in Fig. \ref{domedome6}(a), the density focusing at the center occurs faster than that for $a_{12}=0$, while in region (c) it occurs slower than that for $a_{12}=0$ [see Fig \ref{domedome4}(a)]. This difference is caused by the effect of the intercomponent interaction; the mutual attraction quickens the focusing process of the $\psi_{2}$ component, whereas the mutual repulsion delays the focusing. Another feature is that the dynamic behavior of the dark solitons in $\psi_{1}$ are independent of the behavior of the bright solitons in $\psi_{2}$. These solitons do not form composite vector solitons, because the mutual attraction acts against the coupling of the density dips and the density peaks.

\subsubsection{Condensate dynamics after turning off the trapping potentials}
\begin{figure}
\includegraphics[height=0.24\textheight]{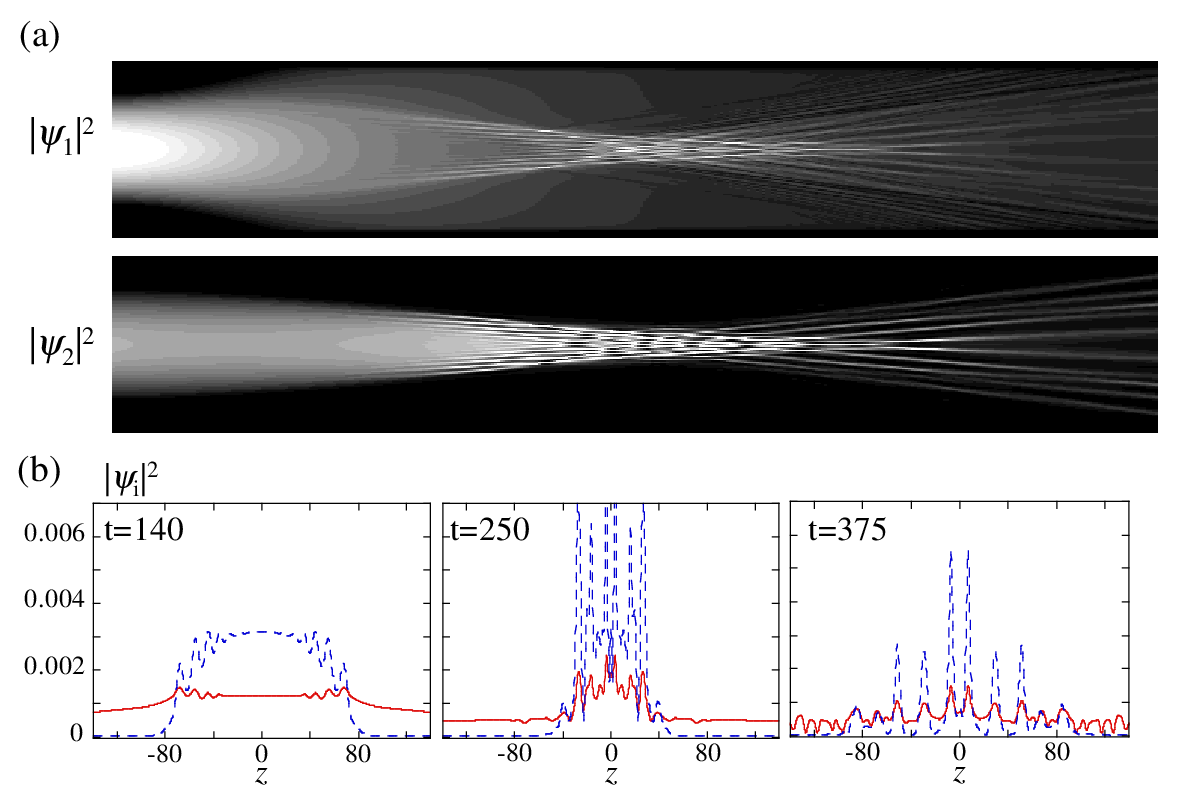}
\caption{(Color online.) Simulation for $N=5 \times 10^{4}$ and $a_{12}=-1.0$ nm, in the case without a trapping potential. (a) The contour plots of the density of both components with respect to time ([0, 500], horizontal axis) and $z$ ([-200, 200], vertical axis). (c) The density profiles of $|\psi_{1}|^{2}$ (solid-curve) and $|\psi_{2}|^{2}$ (dashed-curve) at $t=140$, $t=250$, and $t=375$. The unit of time is $\omega_{\perp}^{-1}$. }
\label{domedome7}
\end{figure}
To see the MI-induced dynamics more clearly, we consider a simple situation in which the two components freely expand by turning off the trapping potential. We first prepared the same initial condition as in the previous section (\ref{aridesu}) and then turned off the trapping potential at $t=0$. Figure \ref{domedome7} represents the results for $N=5 \times 10^{4}$ and $a_{12}=-1.0$ nm. Because of the mutual attraction, the repulsive component expands more slowly than that with the simulation with $a_{12}=0$. After that, the MI starts from the edge of the attractive component and the modulation becomes in-phase for each component. Then, a train of composite solitons forms through the MI. Each soliton is largely the bright component of $\psi_{2}$ with a small fraction of the $\psi_{1}$ component being trapped by the bright soliton. The trapped $\psi_{1}$ component creates the density peaks upon the bright soliton despite the repulsive interaction and the peaks persist during the free expansion. This soliton has been called a bright-antidark solitons \cite{Kevrekidis}. With increasing $|a_{12}|$, the expansion becomes slower and a larger number of sharp composite solitons are formed. A similar trapping mechanism of the solitons is seen in a mixture of bosons and fermions \cite{Karpiuk}; however, the density expansion of the fermions in that case is caused by the Pauli exclusion principle.

\section{CONCLUSION} \label{condle}
We analyzed the modulation instability (MI) and the nonlinear dynamics of multiple solitary-wave formation in trapped two-component BECs. The MI of this system was classified according to the signs and magnitudes of the s-wave scattering lengths. Then, we used the one-dimensional coupled GP equations with the particle loss term to numerically simulate the nonlinear dynamical stage after the MI occurs. For each combination of the scattering lengths, an unstable modulation grew up to a train of vector solitons unique to the two component system. As a result, we obtained the following picture:

(a) When all coupling constants were repulsive, the strong intercomponent interaction caused the phase separation of the two components. The MI first grew near the edge of the condensate, giving rise to solitary waves with alternating condensate domains. These phenomena reproduced the experimental observation by Miesner {\it et al.} \cite{Miesner}. Because the density modulation developed out-of-phase, the total density hardly changed during the domain formation, and this allowed us to reduce the system to a single-component condensate with attractive interactions. 

(b) When the coupling resulted in strongly attractive intercomponent forces, the two components underwent a focusing collapse despite of their repulsive intracomponent interactions. If the condensates were moved to an expulsive potential, the instability of the in-phase modulation generated a vector bright soliton train. In this case, the fact that the two components always overlapped reduced the system to that of a single-component condensate. Thus, the dynamical feature was similar to that in a single-component case \cite{Carr}. 

(c) When one of the components had an intracomponent attractive interaction, the presence of the other component increased the growth rate of the MI over that of the single-component case. Moreover, the unstable dynamics was sensitive to the shape of the trapping potential. For a harmonic potential, the density of the attractive component focused in one spatial region, and this focusing strongly perturbed the repulsive component in this region, which subsequently produced a train of dark solitons. Then, the attractive component coupled with the dark solitons such that the total condensates formed dark-bright solitons. When the trapping potential was different from each other, the MI occurred from the density edge of an initially miscible condensate, leading to the formation of a train of dark-bright solitons. In this case, the formation dynamics was greatly influenced by the intercomponent repulsion. In particular, an increase in $a_{12}$ increased the strength of the MI and increased the number of the solitons. 

(d) When the intercomponent interaction was attractive, the dynamics became more complex. In a harmonic potential, the focusing collapse generated dark solitons in the repulsive component and bright solitons in the attractive one. However, these solitons could not be coupled through the intercomponent attraction. When the trapping potential was turned off, the in-phase modulation developed composite solitons in which some fractions of the repulsive component were trapped by the bright solitons in the attractive component.

Finally we discuss the feasibility of observing the above results experimentally. The situation is that in which two components initially have the same density profile with an inverted parabola and they also have the same position. This can be achieved experimentally by using atoms of the same species but different hyperfine levels. Then, with an rf-pulse, one can instantaneously transfer half of the condensed atoms in one hyperfine level to the other level \cite{Hall,Miesner}. In addition, the Feshbach resonance during atomic collisions depends on both the hyperfine level and the magnetic field. Thus, a suitable choice of the atomic hyperfine levels and control of a magnetic field can realize the parameter regime in Table \ref{coupclas}.

Another way to control the MI condition for two-component BECs is to use a periodic potential \cite{Rapti}, that can change the effective atomic mass. In particular, if the atomic interaction is repulsive, the negative effective mass corresponds to the anomalous diffraction regime and, as a result, the system is modulationally unstable. 

\begin{acknowledgments}
M.T. acknowledges support from a Grant-in-Aid for Scientific Research (Grants No. 15341022) by the Japan Society for the Promotion of Science. 
\end{acknowledgments}


\end{document}